\documentclass[fleqn,usenatbib]{mnras}

\usepackage{newtxtext,newtxmath}
\usepackage{fix-cm}
\usepackage[T1]{fontenc}

\usepackage{float}
\usepackage{graphicx}	%
\usepackage{amsmath}	%
\usepackage{dirtytalk}
\usepackage{siunitx}
\usepackage[dvipsnames]{xcolor}

\DeclareSIUnit{\hred}{\textit{h}}
\DeclareSIUnit{\Msun}{M_\odot}

\usepackage{cleveref}
\crefname{equation}{Eq.}{Eqs.}
\crefname{figure}{Fig.}{Figs.}
\crefname{table}{Table}{Tables}
\crefname{section}{Section}{Sections}
\crefname{appendix}{Appendix}{Appendices}
\Crefname{equation}{Eq.}{Eqs.}
\Crefname{figure}{Fig.}{Figs.}
\Crefname{table}{Table}{Tables}
\Crefname{section}{Section}{Sections}
\Crefname{appendix}{Appendix}{Appendices}

\newcommand{\correction}[1]{\textcolor{black}{#1}}

\usepackage[normalem]{ulem}
\usepackage{enumitem}

\title[Pop III stars in MEGATRON]{MEGATRON:~The environments of Population III stars at Cosmic Dawn and their connection to present day galaxies}

\author[A. Storck et al.]{
Anatole Storck,$^{1}$\thanks{E-mail: storckanatole@gmail.com}
Harley Katz,$^{2,3}$
Julien Devriendt,$^{1}$
Adrianne Slyz,$^{1}$
Corentin Cadiou,$^{4}$
\newauthor{
Nicholas Choustikov,$^{1}$
Martin P. Rey,$^{5}$
Aayush Saxena,$^{1}$
Oscar Agertz,$^{6}$ and
Taysun Kimm$^{7}$
}\\
$^{1}$Sub-department of Astrophysics, University of Oxford, DWB, Keble Road, Oxford OX1 3RH, UK\\
$^{2}$Department of Astronomy \& Astrophysics, University of Chicago, 5640 S Ellis Avenue, Chicago, IL 60637, USA\\
$^{3}$Kavli Institute for Cosmological Physics, University of Chicago, Chicago IL 60637, USA\\
$^{4}$Institut d'Astrophysique de Paris, Sorbonne Université, CNRS, UMR 7095, 98 bis bd Arago, 75014 Paris, France\\
$^{5}$Department of Physics, University of Bath, Claverton Down, Bath, BA2 7AY, UK\\
$^{6}$Division of Astrophysics, Department of Physics, Lund University, Box 118, SE-221 00 Lund, Sweden\\
$^{7}$Department of Astronomy, Yonsei University, 50 Yonsei-ro, Seodaemun-gu, Seoul 03722, Republic of Korea
}

\date{Accepted XXX.\@ Received YYY;\@ in original form ZZZ}

\pubyear{2025}

\begin{document}\label{firstpage}
\pagerange{\pageref{firstpage}--\pageref{lastpage}}
\maketitle

\begin{abstract}
We present results of Population III (Pop III) formation in the \textsc{megatron} suite of simulations, which self-consistently follows radiation and non-equilibrium chemistry, and resolves gas at near-pc resolution in a Milky Way-mass progenitor at Cosmic Dawn. While the very first Pop~III stars form in halos with masses well below the atomic cooling limit, the majority of Pop~III stars form in more massive systems above the $10^4$~K atomic cooling threshold as a Lyman-Werner (LW) background of $10^{-21}\,\rm erg\,s^{-1}\,cm^{-2}\,Hz^{-1}\,sr^{-1}$ is rapidly established. We find that the global Pop~III star formation rate stabilizes to a value of $10^{-3}\,\rm M_\odot\,yr^{-1}$ at $z=20$. Among the three processes that quench Pop~III star formation in mini-halos, the LW background, gas starvation, and external chemical enrichment, the LW background is most important. A small fraction of halos undergo multiple episodes of Pop~III star formation when the earlier forming stars all directly collapse to black holes. If the halos become massive enough, they can form up to $\sim100$ Pop~III stars in a single burst, which may be observable by JWST with moderate gravitational lensing. Pop~III stars form at a wide range of distances from UV-bright galaxies, with only $0.06\,$\% of Pop~III stars forming within the virial radius of galaxies with $M_{\rm UV}<-17$. Finally, by tracking Pop~III star remnants down to $z=0$, we find that $75-80\,$\% reside in the stellar halo of our simulated Milky Way analogue, while the remainder are gravitationally bound to lower-mass systems, including satellite halos.
\end{abstract}


\section{Introduction}\label{seq:Introduction}

Big Bang nucleosynthesis (BBN) predicts that gas in the Universe was initially composed of hydrogen, helium, and trace amounts of lithium. This metal-free `pristine' or `primordial' gas adiabatically cooled and began to gravitationally collapse in overdensities following its decoupling from the radiation field at $z\sim 1100$, and was slowly funnelled into the first dark matter halos. The first stars in our Universe formed from this primordial gas in small and molecular hydrogen-cooling dark matter halos ($\sim 10^5-10^6\;\rm M_\odot$; mini-halos) around $z\sim 30$ \citep{Barkana_Loeb_2001, Greif_2015}. Such metal-free, Population III (Pop~III) stars, synthesized the first metals in their cores and began the process of enriching the Universe with heavy elements \citep{Bromm_Loeb_2003}. The ionizing photons from Pop~III stars began cosmic Reionization and brought about an end to the cosmic Dark Ages \citep{Miralda-Escude_2003}. Most haloes in the early Universe once hosted Pop~III stars, playing a critical role in the initial evolution of galaxies \citep{Johnson_Greif_Bromm_2008,Bromm_2009}. Consequently, Pop~III stars are a crucial component for our understanding of the Universe at Cosmic Dawn.

Pop~III stars have theoretically different structures than later-time stellar populations. Their atmospheres are not subject to metal-driven cooling instabilities, which produce stellar winds \citep{Vink_2000}, unless metals cycle throughout \textit{via} mixing in fast rotators \citep{Ekström_2008}, and their cores do not have enough carbon to ignite the CNO cycle in the initial hydrogen-burning stage. Observationally, little is known about them, and theoretically, simulations still disagree on their properties. The masses of Pop~III stars are notoriously uncertain, with estimates for the Pop~III Initial Mass Function (IMF) differing by orders of magnitude in characteristic mass, and profiles ranging from uniform to log-normal \citep[e.g.][]{Hirano_Hosokawa_Yoshida_Umeda_Omukai_Chiaki_Yorke_2014, Wollenberg_Glover_Clark_Klessen_2020, Jaura_Glover_Wollenberg_Klessen_Geen_Haemmerlé_2022}. Constraining the Pop~III IMF is crucial for understanding the dominant feedback mechanisms of the first stars, whose metals brought forth a new generation of enriched, Pop~II stars, and regulated the growth of the earliest galaxies \citep{Smith_Wise_OShea_Norman_Khochfar_2015,Jaacks_2018}. Although estimates of the mass distribution of Pop~III stars vary, it is generally accepted that the Pop~III IMF is top-heavy compared to present-day stellar populations. The best constraints come from the non-detection of Pop~III stars in the stellar halo of the Milky Way \citep[e.g. with SDSS,][]{Yanny_SDSS_2009}, and from the chemical signatures of local metal-poor stars \citep{Ishigaki_2018,Jiang_2024}.

The majority of Pop~III stars likely form in isolation or within small ($\leq 10^4\,\rm M_\odot$) stellar clusters at high redshift \citep{Bromm_Yoshida_2011}, making them extremely difficult to observe directly even with our most sensitive detectors. As a consequence, there has not yet been a galaxy observed with a confirmed detection of a Pop~III star at high-$z$, although many Pop~III galaxy candidates have been proposed \citep{Vanzella_NIRSpec_2023,Maiolino_JADES_2024,Wang_NIRSpec_2024,Cai_JWST_VLT_2025,Fujimoto_glimpse_2025,Nakajima_2025,Vanzella_2025}. Gravitational lensing would help resolve individual Pop~III stars if the magnification is large enough. Although the probability of such a configuration is low, we are already finding metal poor stars/clusters in highly magnified ($\mu\sim1000$) caustics \citep{Welch_2022}. If Pop~III stars can form with masses down to $\sim 0.8\,\rm M_\odot$, they may survive to the present day \citep{Maeder_Meynet_Pop3Review_2012}, and be identified in our local group of galaxies from their chemical signatures \citep{Tanaka_Chiaki_Tominaga_Susa_2017}. So far, no Pop~III stars have been found in archaeological surveys of the Milky Way or local dwarfs, which places a constraint on the lower bound of the Pop~III IMF to be $\sim0.7-0.8 \,\rm M_\odot$ \citep{Salvadori_2010,Hartwig_Bromm_Klessen_Glover_2015}.

The study of Pop~III stars has remained primarily theoretical due to the difficulties of directly observing them or their remnants. With more sensitive telescopes, there is a need for numerical simulations and semi-analytical models to make predictions of galaxies hosting Pop~III systems, such as predictions of their unique spectral features \citep[e.g.][]{katz_pop3_2023,Trussler_2023}, how long the Pop~III phase lasts \citep[e.g.][]{Xu_pop3_2016,Hegde_Furlanetto_2025}, and the types of environments which stimulate/prevent Pop~III formation \citep[e.g.][]{Park_2021,Nishijima_Hirano_Umeda_2024}. Upcoming spectral surveys of local stars (e.g 4MOST \citealp{Christlieb_2019} and WEAVE \citealt{Jin_2024}) also need better constraints on where Pop~III stars end up at present day to aid in targeted searches of these metal-free relics.

The wide range of spatial scales required to understand the Pop~III formation phase forces simulations of Pop~III stars to typically either focus on the small scale clouds forming hydrostatic cores to predict the Pop~III IMF \citep[e.g.][]{Clark_Glover_Klessen_Bromm_2011,Hirano_Hosokawa_Yoshida_Umeda_Omukai_Chiaki_Yorke_2014,Hosokawa_Hirano_Kuiper_Yorke_Omukai_Yoshida_2016}, or on the large scale galactic context in which they form to produce statistical studies of Pop~III birth sites and their subsequent impact on galaxy formation \citep[e.g.][]{wise_pop3IMF_2012, Xu_pop3_2013, kimm_pop3_2017, katz_pop3_2023, Brauer_AEOS_2025}. Environmental studies have been typically limited to small ($\sim 2\,\mathrm{cMpc}/h$) cosmological boxes that sample a limited number of halos and environmental configurations (although see e.g. the Renaissance simulations \citep{Xu_pop3_2013} for an overdense $\sim 40\,\mathrm{cMpc}^3/h^3$ volume).

The H$_2$ content in halos is a critical component in determining whether halos will eventually form a Pop~III star, with the lower bound on halo mass for Pop~III formation, called $M_\mathrm{crit}$ or \correction{$M_\mathrm{min}$}, being highly dependent on the production/destruction rate of H$_2$. A key regulator in the production of H$_2$ in pristine halo cores is Lyman-Werner (LW) radiation, which travels far from young galaxies and across the inter-galactic medium (IGM), photodissociating the molecular hydrogen and preventing cooling \citep{Haiman_Abel_Rees_2000} up to densities where H$_2$ shielding can become effective. 
Results from the Renaissance simulations \citep{Xu_pop3_2013, Xu_pop3_2016} show that the Pop~III formation phase can continue past the late stages of reionization \citep[see also e.g.][]{Zier_THESAN_2025}, extended from the delay of star formation by LW radiation in halos below the atomic cooling limit \citep[see also e.g.][]{OShea_2008}.
It is still unclear how this LW background, which is expected to be anisotropic and environment-dependent, affects potential Pop~III birth sites in a system like our Milky Way.

To address these scientific questions, we utilize the \textsc{megatron} simulations. \textsc{megatron} is a suite of high-resolution zoom simulations at Cosmic Dawn of the Lagrange region of a halo that has approximately Milky Way mass by $z=0$. The volume of the zoom region is $20\,\mathrm{cMpc}^3/h^3$, which allows for a comprehensive statistical study of Pop~III stars in the context of a Milky Way-like environment. \textsc{megatron} forms individual Pop~III stars, and models the transition to Pop~II stars, which will increasingly contribute to the LW background radiation as the Pop~II star formation rate (SFR) begins to dominate the total SFR. With its near-pc resolution for gas at $z\sim 30$, radiative transfer coupled to a complex non-equilibrium chemistry network including H$_2$, \textsc{megatron} is ideal for studying the formation sites of the first stars in our Universe. \textsc{megatron} includes a dark matter-only simulation run to $z=0$, which we utilize to trace where the remnants of Pop~III stars end up at present day.

This paper is structured as follows. We introduce the \textsc{megatron} suite of simulations in Section~\ref{seq:Method}, including our treatment of Pop~III star formation. We then present our results in Section~\ref{seq:Results}, and conclusions in Section~\ref{seq:Conclusions}.
\section{Method}\label{seq:Method}

\textsc{megatron} is a suite of cosmological radiation hydrodynamic simulations run with the \textsc{ramses-rtz} code \citep{katz_ramses-rtz_2022} and the PRISM ISM model \citep{katz_prism_2022}, extended from \textsc{ramses-rt} \citep{rosdahl_radiative_2013, rosdahl_scheme_2015} and \textsc{ramses} \citep{teyssier_ramses_2002}. The simulations model gravity, hydrodynamics coupled to multi-frequency radiative transfer, and a non-equilibrium chemistry network of over 80 species to accurately model heating and cooling processes. In this section, we give a brief overview of the four high-$z$ simulations used in our analysis, \correction{which have been run down to $z=8.5$}, focusing on the details which are important in the context of Pop~III star formation. For details on how we treat particle dynamics, gas dynamics, and radiative transfer, see \citet{Katz2025MegP1}.

\subsection{Initial conditions \& Resolution}

Initial Conditions (ICs) are generated with \texttt{genetIC} \citep{stopyra_genetic_2021,pontzen_genetic_2022} using cosmological parameters from the Planck data release (\citet{planck_data_2016}; $\Omega_{\rm m}=0.3139$, $\Omega_\Lambda=0.6861$, $\sigma_8=0.8440$, $n_{\rm s}=0.9645$, $h=0.6727$) in a ${(\SI{50}{cMpc/\hred})}^3$ box. The ICs are originally taken from \citet{Rey_Starkenburg_2022} and modified following the procedure in \citet{pontzen_nodifussion_2021} to mitigate gas advection errors. We apply quadratic modifications \citep{rey_quadmod_2018} to the density field, producing an early forming scenario for a MW progenitor and yielding a large number of haloes at high-$z$ that are bright enough to be observed by JWST. \correction{Baryon-dark matter streaming velocities are not included in the ICs.}

As with most zoom simulations, an initial large volume dark matter-only simulation is run to identify a Milky Way-mass ($\sim 10^{12}\,\rm M_\odot$) halo at $z=0$. All particles within $3\,r_{\rm vir}$ are then traced back to the initial conditions, defining the Lagrangian patch. The simulation is then re-centred on the $z=0$ position of the halo of interest, and a zoom region is constructed as a set of nested grids covering the Lagrangian patch, reaching a maximum mass resolution of $\SI{2.48e4}{\Msun/\hred}$ for the dark matter (an effective resolution of $8196^3$ in a $\SI{50}{cMpc/\hred}$ box). The zoom region encompasses a volume of ${\left(2.7\;{\rm cMpc}/h \right)}^3\approxeq 20\,\mathrm{cMpc}^3/h^3$.

Refinements of the grid are performed according to the local fluid properties (total mass in the cell, jeans length) down to $\SI{48}{cpc/\hred}$. Since the refinement criteria is in co-moving coordinates, the physical resolution is initially much smaller than $\SI{48}{pc/\hred}$, reaching $\SI{1.5}{pc/\hred}$ resolution at the onset of Pop~III formation $(z\sim30)$. This high resolution combined with our top-heavy Pop~III IMF (described later on), means that almost all Pop~III stars have resolved Strömgren spheres in the initial stages of Pop~III formation.

\subsection{Chemistry \& Radiation}

\correction{The high-$z$ suite of \textsc{megatron} simulations} are initialized with the primordial composition predicted from BBN, allowing the metallicity to grow naturally from pollution by Pop~III and Pop~II stars.
Metallic species are followed in non-equilibrium if they contribute at least 1\% to the equilibrium cooling curve derived from \textsc{cloudy} \citep{ferland_cloudy_2017} across the temperature range $10^1-10^8\,\rm K$. We track primordial species: $\mathrm{e^-}$, H\,\textsc{i-ii}, He\,\textsc{i-iii};
 molecules: H$_2$ %
 and CO; %
and the following metallic ions: O\,\textsc{i-viii}, C\,\textsc{i-vi}, N\,\textsc{i-vii}, Fe\,\textsc{i-xi}, Si\,\textsc{i-xi}, S\,\textsc{i-xi}, Ne\,\textsc{i-x}, and Mg\,\textsc{i-x}. Specifically, ions can be created or destroyed through ionization, recombination, and charge exchange processes following \citet{katz_prism_2022}, or by forming molecules. See \citet{Katz2025MegP1} for how dust is modelled.
The net cooling rate $\Lambda_{\rm net}$ is computed from primordial species%
, H$_2$%
, CO%
, dust processes%
, and metallic transitions. If a specific metallic ion is not directly tracked, its contribution to the cooling curve is still considered under the assumption of collisional ionization equilibrium. The net heating rate $\Gamma_{\rm net}$ is computed from photoheating from stellar SEDs, photoelectric heating from dust%
, and H$_2$ heating %
(formation, dissociation, pumping.)

\correction{
    The radiation from stars is propagated in 8 frequency bins covering the infrared (IR) to the extreme ultraviolet (EUV), transferring energy by photoionization (EUV-Opt) and photoheating (FUV) of the gas, photodissociation of molecular hydrogen (LW), and momentum through radiation pressure (Optical-IR). Further details on the method of radiative transfer can be found in \citet{Katz2025MegP1}.
    The LW background from sources outside the zoom region is considered to be sub-dominant compared to the background generated by our early collapsing region, and the ionization flux from the included UV background \citep{Haardt_Madau_2012} is negligible at the redshifts we consider.
}

\subsection{Star formation in the fiducial model}\label{subseq: SF fiducial}

\correction{The high-$z$ suite of \textsc{megatron} simulations} have various models for star formation. In this section, we describe the fiducial model which we call the {\bf Efficient SF} model, and describe star formation in alternative models in the next section. Star formation proceeds according to the turbulent properties of the gas, and is similar to the SF model used in the SPHINX \citep{Rosdahl_SPHINX_2018} simulation. A gas cell is able to form stars if its gas density is $>10\;m_{\rm p}\;\rm cm^{-3}$ and $>200\times\rho_\mathrm{crit}$ with $\rho_\mathrm{crit}$ the critical density, the turbulent Jeans length is smaller than the cell size, the gas flow is convergent across the adjacent cells, and the gas density field is at a local maxima. The efficiency of star formation $\epsilon_{\rm ff}$ is computed from the local turbulence (\citealt{Padoan_2011,Federrath_2012}, see \citealp{katz_impact_2024}). For Pop~III star formation each star particle represents an individual star, while Pop~II stars are sampled as simple stellar populations.

\begin{figure}
	\includegraphics[width=0.47\textwidth]{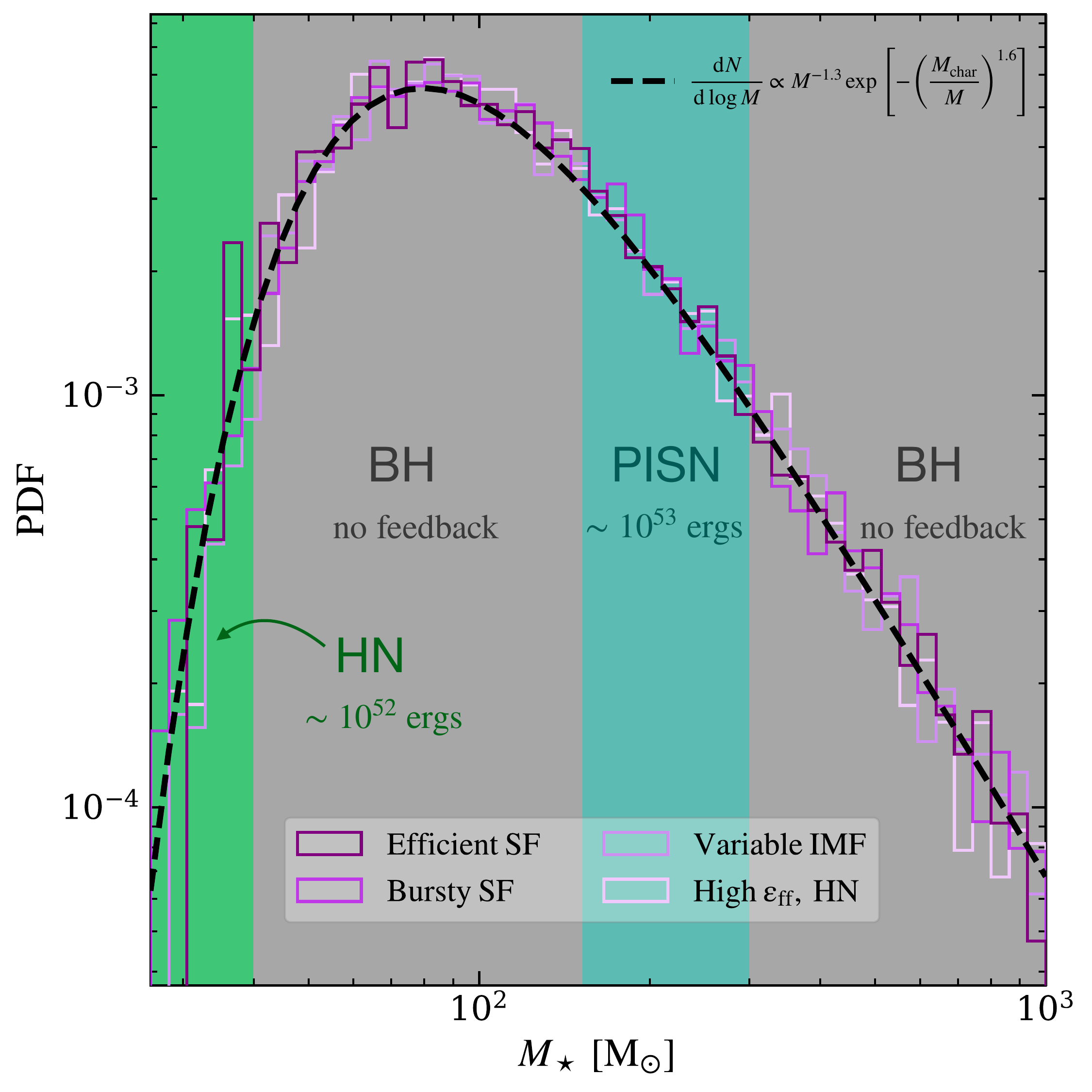}
    \caption{
            The Initial Mass Function (IMF) of Pop~III stars used in the \textsc{megatron} suite of simulations. Histograms of Pop~III masses for each run are included to show they all sample the log-normal IMF. Regions in the background are coloured according to the feedback channels a star of particular initial mass will follow, starting with hypernova (HN) and moving to black hole (BH) formation with a narrow mass regime where the star can undergo pair-instability supernova (PISN). The Pop~III model also includes CCSN at lower masses. They are however probabilistically unlikely to be sampled due to the chosen characteristic mass of $\sim 100\,\rm M_\odot$.
    }\label{fig:pop3_IMF}
\end{figure}

\subsubsection{Pop~II model}\label{subseq: pop2 model}

At metallicity $Z > 2\times10^{-8}$ (the transition metallicity between Pop~II and Pop~III stars, where gas cloud fragmentation becomes more efficient \citealt{Omukai_2005}), we probabilistically sample the formation of star particles in integer multiples of $500\,\rm M_\odot$ based on a Poisson distribution. Stellar SEDs are computed based on the metallicity and age of star particles using BPASSv2.2.1 \citep{Eldridge_BPASS_2008, Eldridge_BPASS2p1_2017, Stanway_BPASS2p2_Eldridge_2018}, and scaled by mass of the star particle. Similarly, CCSN are stochastically sampled at each time-step using the BPASSv2.2.1 IMF, where CCSN progenitors are stars with initial masses ranging from $8-25\;\rm M_\odot$. They release $10^{51}\;$ergs of energy at the time of the explosion. For details on metal yields and stellar winds of Pop~II stars \correction{in \textsc{megatron}}, see \citet{Katz2025MegP1}.

\subsubsection{Pop~III model}\label{subseq: pop3 formation}

Below metallicity $Z = 2\times10^{-8}$, individual Pop~III stars form from the following the IMF
\begin{equation}
    \frac{\mathrm{d}N}{\mathrm{d}\log M} \propto M^{-1.3} \exp \left[-{\left(\frac{M_{\rm char}}{M}\right)}^{1.6}\right],
\end{equation}
with a characteristic mass $M_{\rm char} = \SI{100}{\Msun}$ \citep{wise_pop3IMF_2012}.

We assume that once a Pop~III star is formed, it remains on the main sequence with no change to its luminosity or mass until hydrogen burning ends. Stellar SEDs for Pop~III stars are interpolated in age and mass following tabulated values from \citet{schaerer_pop3sed_2002}.

\begin{figure}
	\includegraphics[width=0.47\textwidth]{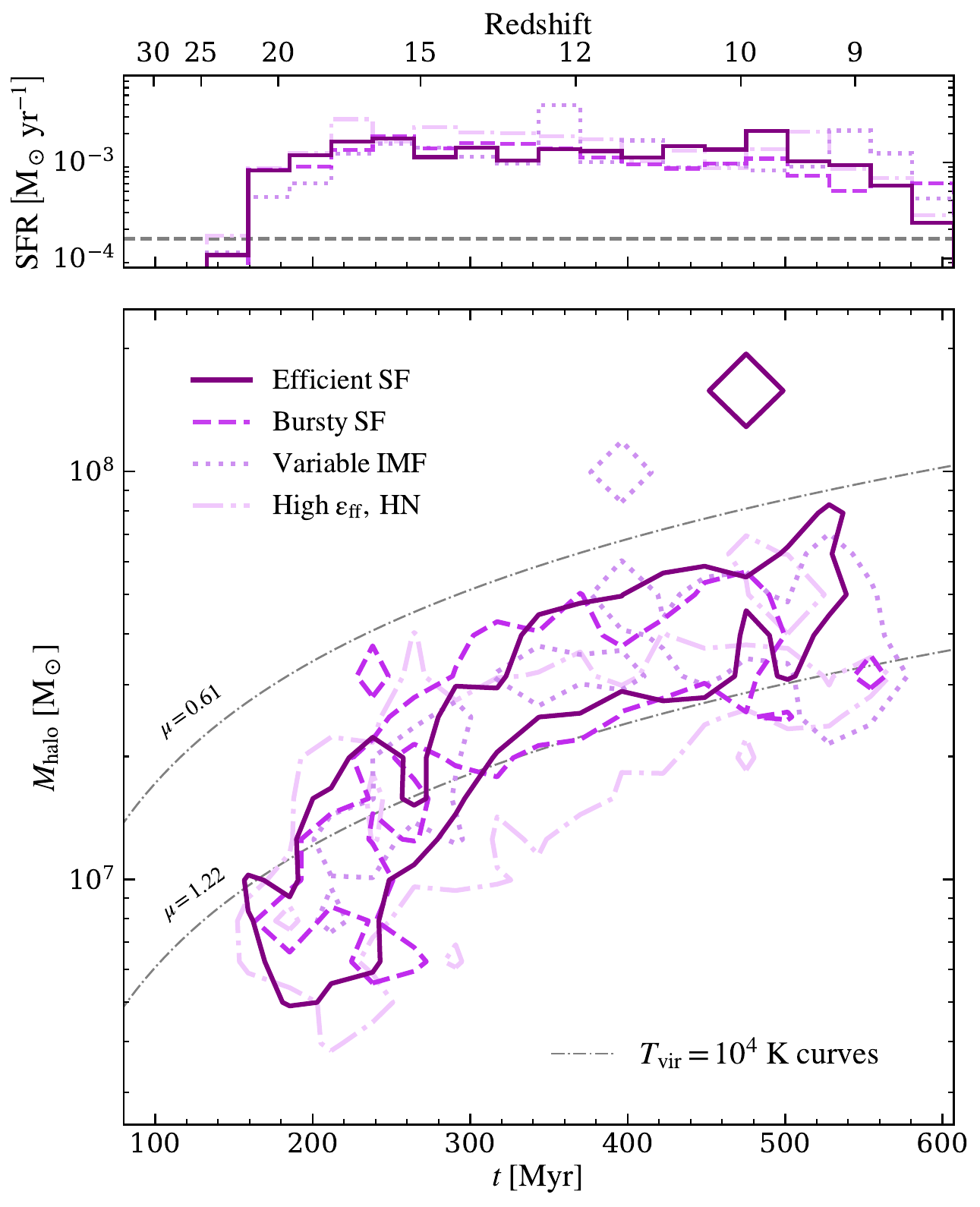}
    \caption{
            \textbf{(Top)} Star formation rate of Pop~III stars as a function of time from the start of the simulations. The SFR quickly rises to $10^{-3}\,\rm M_\odot\,yr^{-1}$ and remains mostly constant for all simulations until the end of the simulation at $z=8.5$. The gray dashed line at $\sim 1.7\times10^{-4}\;\rm M_\odot \,yr^{-1}$ is the level for the contour in the lower panel.\ \textbf{(Bottom)} Contours in halo mass and time of halos hosting Pop~III stars weighed by the SFR of Pop~III stars, for each simulation. The mass-time tracks for halos with a virial temperature of $10^4\;$K \citep{Barkana_Loeb_2001} are shown as the gray dash-dotted lines, labelled by mean molecular weight ($\mu=1.22$:~neutral gas; $\mu=0.61$:~ionized gas). For all simulations, Pop~III formation migrates from mini-halos to atomic cooling halos on a short timescale. Pop~III starbursts in $10^8\,\rm M_\odot$ halos are visible for the Efficient SF and Variable IMF runs.
    }\label{fig:pop3_masstime_SFR}
\end{figure}

At the end of their lives, Pop~III stars follow four different mass-dependent phases, three of which we show in Fig~\ref{fig:pop3_IMF}. Stars between $10-20\,\rm M_\odot$ undergo normal core-collapse supernova (CCSN) releasing $\SI{e51}{ergs}$.\footnote{Due to the low sampling probability, our simulations contain no stars with low enough masses for CCSN.} At slightly higher masses, stars between $20-40\,\rm M_\odot$ explode as hypernova (HN) injecting energies of $10-40\times 10^{51}\;\rm ergs$ depending on mass \citep{Nomoto_Yields_2006}. From $40-140\,\rm M_\odot$, and above $300\,\rm M_\odot$, we assume direct collapse to a black \correction{hole} with no supernova explosion. In the intermediate mass range, stars between $140-300\,\rm M_\odot$ explode as pair-instability supernovae (PISN), releasing energy depending on the remaining He core mass \citep{Heger_PISNHeCore_2002}.
The above feedback mechanisms from Pop~III stars will enrich the surrounding medium with metals, with differing chemical yields depending on the type of supernova. We adopt chemical yields for the various feedback processes from \citet{Umeda_Yields_2002}; \citet{Nomoto_Yields_2006,Nomoto_Yields_2013}.

\subsection{Star formation in alternative models}\label{subseq: SF alternatives}

We introduce here three additional simulations which have varying physics for Pop~II star formation, to explore the robustness of our results to changes in galaxy formation physics. The {\bf Bursty SF} model increases the energy of Pop~II supernovae from $10^{51}\;\rm ergs$ to $5\times10^{51}\;\rm ergs$. The {\bf Variable IMF} model allows for a Pop~II IMF which is dependent on the local gas density and metallicity computed using Starburst99 \citep{Leitherer_Starburst_1999}, along with adding hypernova for Pop~II stars. Finally, the {\bf HN + High $\epsilon_{\rm ff}$} model combines Pop~II hypernova feedback with a stricter jeans length criterion of four times the grid resolution, and uses a constant $\epsilon_{\rm ff} = 100\%$. Hypernovae in the last two models occur for Pop~II stars above $25\;\rm M_\odot$. The Pop~III model is identical in all runs with the only difference being the Jeans criteria for star formation in the {\bf HN + High $\epsilon_{\rm ff}$} model.

\subsection{Generating spectra}\label{subseq: generating spectra}

Galaxy spectra are generated from gas cells and star particles, located within $0.25\,r_{\rm vir}$ of the halo centre, by summing the stellar SEDs from both Pop~II and Pop~III stars, the nebular emission, and the nebular continuum.
Nebular emission of \correction{recombination and collisionally excited lines} are computed using \textsc{pyneb} \citep{Luridiana_pyneb_2015} and the CHIANTI database \citep{Dere_chianti_2019}, where we build up emissivity grids in temperature and electron density for each line. We correct for emission lines from cells with unresolved Strömgren spheres using CLOUDY (\citealt{ferland_cloudy_2017}, see \citealt{Katz2025MegP1} for details), and the nebular continuum is computed using \textsc{pyneb}.
When processing the Pop~III SEDs for observability, we use mass-dependent SEDs derived from Pop~III atmospheric modelling \citep{larkin_pop3atmos_2023}.
To compare with observational instruments such as JWST, we convert the spectra from our computed luminosities $L$ to AB magnitudes: 
\begin{equation}
    m_\mathrm{AB} = -2.5\log_{10}\left(\frac{L_\nu}{4\pi D^2_\mathrm{L}}(1 + z)^{-1} \frac{\rm s\,cm^2\,Hz}{\mathrm{erg}}\right) - 48.6,
\end{equation}
where the luminosity is per frequency unit and $D_\mathrm{L}$ is the luminosity distance. We convolve the spectra with selected filters from JWST/NIRCam and JWST/MIRI using \textsc{sedpy} \citep{Johnson_SEDPY_2019}. UV magnitudes of galaxies are computed following \citet{Oke_Gunn_1983} as
\begin{equation}
    M_\mathrm{UV} = 51.595 - 2.5\log_{10}\left(\frac{L_{1500\textup{~\AA}}}{\rm erg \, s^{-1} \, Hz^{-1}}\right).
\end{equation}

\subsection{Structure finding}

For the purpose of this work, we use a \correction{different} halo catalogue from that used in the main simulation paper \citep[\correction{which also includes an analysis of Pop~III systems;}][]{Katz2025MegP1}.
We identify dark matter halos and subhalos using the \textsc{rockstar-galaxies} structure finder \citep{behroozi_rockstar_2013}, retaining all halos with more than 30 particles ($\gtrsim \SI{e6}{\Msun}$) to ensure completeness on halos forming Pop~III stars, down to halo masses of $7\times10^5\,\rm M_\odot$. Halo merger trees are constructed using the \textsc{yt} \citep{turk_yt_2011} and \textsc{tangos} \citep{pontzen_tangos_2018, pontzen_tangos_2023} packages. We make use of the \texttt{crosslink} command from \textsc{tangos} to link the same DM structures across the different \textsc{megatron} simulations if they are initialized with the same ICs, allowing us to track the same halos across different runs. 

\begin{figure*}
	\includegraphics[width=0.99\textwidth]{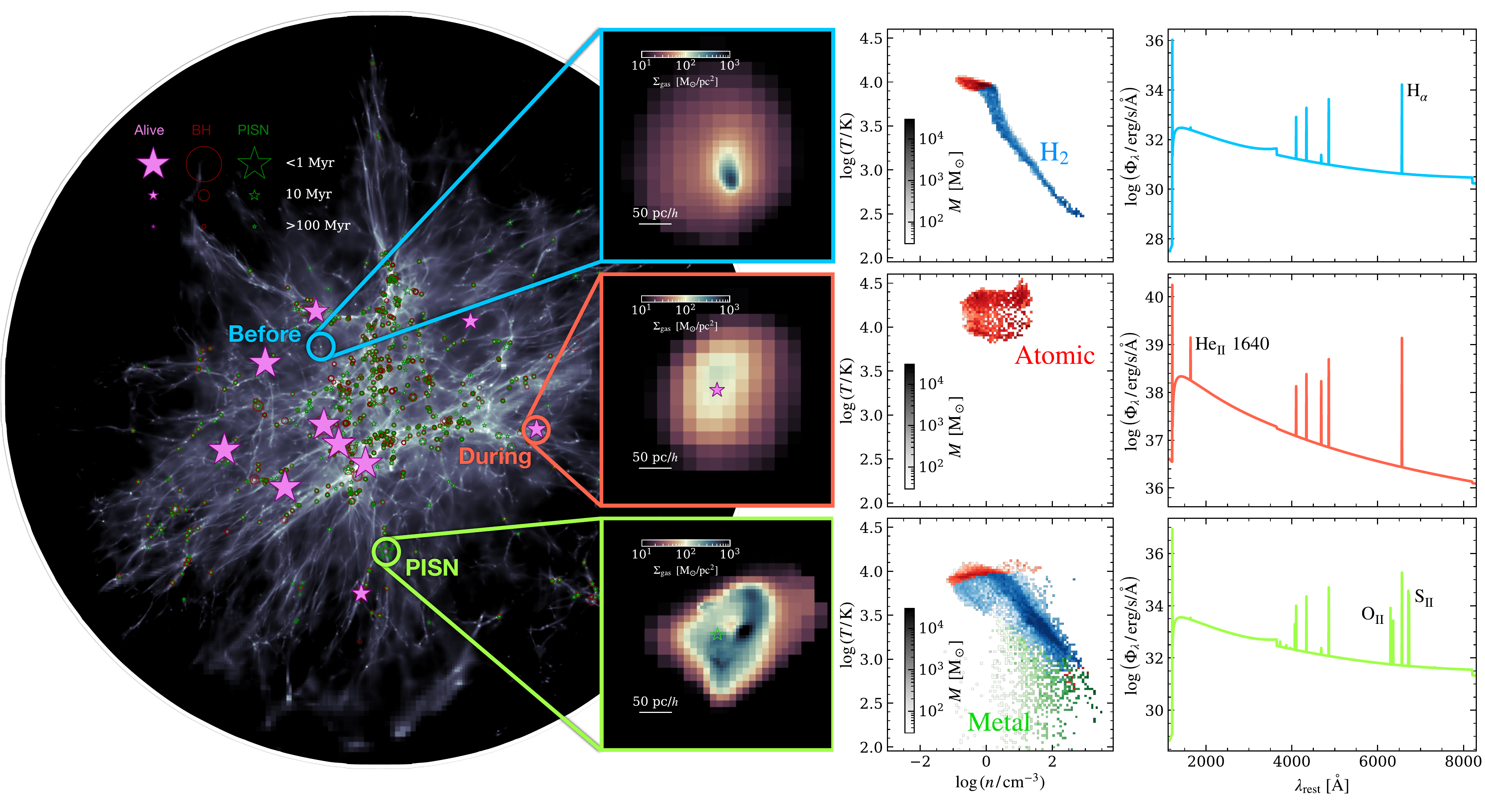}
    \caption{
    \textbf{(left)} Projection of the gas density of the Lagrangian volume in the {\bf Efficient SF} simulation at $z=8.5$. All Pop~III stars which have formed by that redshift are over-plotted on the projection, with symbol sizes corresponding to their age, and colours/markers indicating if the star is still alive or a remnant. Most of the stars older than $5\;\rm Myr$ are now remnants, either a black hole or pair-instability supernova, and are mostly concentrated in the high-density filamentary structure, compared to young Pop~III stars which live on the outskirts. In the insets of the figure, we highlight three different halos displaying the three stages of Pop~III formation: before the star forms, while the star is alive, and the resulting PISN if the star meets the initial mass requirement $(140 < M_\star < 300)$. \textbf{(right)} Each row is for a particular stage of Pop~III evolution. The first column is the gas density projection in the inner part of the halo, the second column is the halo's phase diagram weighed by mass and coloured by dominant cooling process (blue: H$_{2}$, red: atomic, green: metal cooling), and the third column is the halo's intrinsic spectrum built up from nebular emission and continuum (and for the middle row the Pop~III SED.)
    }\label{fig:pop3_stages}
\end{figure*}

\subsection{Tracing Pop~III stars}\label{subseq: pop3 traceback method}

Running the simulations to $z=0$ is too computationally expensive. To find where Pop~III stars reside at $z=0$, we trace the positions of Pop~III stars in the simulations at the last available snapshot $(z\sim 8.5)$ to $z=0$ using a dark matter-only (DMO) simulation run with the same initial conditions and the same DM mass resolution. The large scale formation of structures is driven by the dark matter field, and should proceed to form the same halos, at least in the regime of the Pop~III formation phase. While non-linear coupling from baryonic feedback will cause the matter-power spectrum to deviate from that of a pure DM result at low-$z$, we expect mergers and tidal stripping between halos in the DMO run to follow closely that of a hydro run. 

We match the IDs of DM particles that are near Pop~III remnants at $z=8.5$ with their positions in the DMO simulations at $z=0$. In our analysis, we trace groups of DM particles using a physical distance criteria from the Pop~III star. %
For the purpose of this work, we choose to track the 25 DM particles which are closest physically to each Pop~III star at $z\sim 8.5$. We find them at $z=0$ in the DMO simulation and assign them to the main halo or subhalos according to the halo catalogue. 

\correction{
    We have validated this method by using the full hydrodynamic simulation, tracing early forming Pop~III stars ($z>20$) to the end of the simulation ($z=8.5$) using their neighbouring dark matter and comparing to the true position. The majority of stars are traced to within 200 pc of their true position.
}
\correction{
    Furthermore, in the DMO simulation, we find that for halos which survive to $z=0$, the most bound particles at $z=8.5$ remain bound to the halo at $z=0$ (see Appendix~\ref{app:traceback_validation}). Pop~III stars typically form close to the gravitational centres of their host halos. If the stars reside in subhalos that remain bound and relatively undisturbed, we expect our method to be robust in identifying their $z=0$ hosts. However, if the stars initially reside in halos that merge into the main halo, we expect the stars to have virialized with the main halo by $z=0$.
}

\section{Results}\label{seq:Results}

In this section, we present our results on the formation of Pop~III stars in the \textsc{megatron} suite of simulations. The simulations form \correction{[2655, 2392, 2721, 3217]} Pop~III stars by $z=8.5$ for the {\bf Efficient SF}, {\bf Bursty SF}, {\bf Variable IMF}, and {\bf HN + High $\epsilon_\mathrm{ff}$} runs respectively. As we will discuss later, there can be many repeated events of Pop~III formation in the same halo, so we also consider the number of unique birth sites which produce Pop~III stars. The simulations contain \correction{[882, 858, 863, 1180]} unique halos hosting Pop~III formation. In comparison, \citet{Griffen_2018} found $\sim550$ Pop~III-forming sites for a Milky Way progenitor (when also accounting for radiative feedback), although their semi-analytic model does not self-consistently follow the LW background.

We plot the evolution of the Pop~III halo mass regime as a function of time in Figure~\ref{fig:pop3_masstime_SFR}. The initial formation of Pop~III stars at around $z=30$ is facilitated by molecular hydrogen cooling in $\sim3\times10^6\,\rm M_\odot$ halos, which is quickly shut down by Lyman Werner (LW;\ 11.2–13.6~eV) radiation from the first stars. While not reflective of the average, some Pop~III stars are able to form in halos slightly below $10^6\,\rm M_\odot$. Pop~III formation transitions to occurring in more massive halos, above the atomic cooling limit, as the background radiation field increases. Atomic cooling, primarily facilitated by the Lyman-$\alpha$ transition, enables halos to reach high enough densities for H$_2$ self-shielding against LW to become effective. 
\correction{
    Although the formation rate of Pop~III stars in mini-halos is already small compared to those in atomic cooling halos, our results could overestimate the number of Pop~III stars forming in mini-halos as we do not explicitly model the effect of baryon-dark matter streaming velocities \citep{Tseliakhovich2010, Athena2011, Fialkov_Barkana_Tseliakhovich_Hirata_2012, Schauer2021}. Indeed, a large enough local streaming velocity\footnote{\correction{The formation of the MW may have been subject to a particularly high streaming velocity \citep[$v_\mathrm{SV}=1.75\sigma_\mathrm{SV}$;\ ][]{Uysal_Hartwig_2023}.}} acts against gravitational collapse and delays star formation, increasing the critical mass $M_\mathrm{crit}$ at a given redshift.
}

The global Pop~III star formation rate (SFR) quickly rises to $10^{-3}\rm\, M_\odot \, yr^{-1}$ and remains approximately constant until the end of the simulation. A constant Pop~III SFR down to the late stages of reionization is also observed in other simulations \citep[e.g][]{wise_pop3IMF_2012,Xu_pop3_2016}. If we divide the Pop~III SFR by the volume of the Lagrangian patch, our constant global Pop~III SFR translates to a Pop~III SFR density (SFRD) of $\sim 7\times10^{-5}{\rm\, M_\odot \, yr^{-1} \, Mpc^{-3}} \, h^3$. As our environment is overdense and biased towards building up large halos at high redshift, we expect our SFRD to be large compared to an under-dense environment. Many other works predict similar values for the Pop~III SFRD \citep{Hartwig_Magg_2022}, although most exhibit a continuous increase of the SFRD in time until $z\sim 10$ \citep{Visbal_2020, Liu_Bromm_2021,Ventura_2025}, with some reaching an order of magnitude higher SFRD than our work by that time \citep{deSouza_2014, Jaacks_2018}. While our SFR hints at a slow decline near $z=8.5$, such behaviour may not necessarily continue down to lower redshift. For example, \citet{Hegde_Furlanetto_2025} find a decreasing Pop~III SFRD from its peak at $z=15$ until $z=8$ due to the increasing scarcity of Pop~III-forming halos below the atomic cooling limit, at which point the SFRD continuously increases to $z=5$ as atomic cooling halos take over. In our simulations, we find the transition of Pop~III formation from mini-halos to atomic cooling halos occurs almost immediately after the first stars form.

In \textsc{megatron}, we follow the dominant H$_2$ formation channel facilitated by H$^-$ (Equation~\ref{eq:H2}) and do not consider other, sub-dominant mechanisms such as the formation of H$_2$ through H$^+$ or H$^+_2$. Our chemistry network also does not explicitly follow the formation of HD, which allows for cooling below 100~K in metal free gas \citep{Nagakura_HD_2005}. Some studies which include a more comprehensive 11/12-species primordial chemistry network \citep{Lenoble_2024} recover a similar lower mass threshold for Pop~III formation, and higher resolution \citep[both in gas cells and DM particles,][]{wise_pop3IMF_2012, kimm_pop3_2017} recover a similar mass range of halos forming Pop~III stars. Whether we cool exclusively through H$_2$ or also HD, the halo will eventually form Pop~III star(s). HD cooling can lead to lower core temperatures in atomic cooling halos, which could affect the Pop~III IMF through further cloud fragmentation \citep{Bromm_2009}.

\subsection{The environments of Pop~III systems}\label{subseq: pop3 environments}

\subsubsection{Before the Pop~III star forms}\label{subsubseq: pop3 before formation}

The initial state of gas in a halo at the onset of Pop~III formation is pristine, with cooling driven by H$_2$ in the inner regions down to 300~K. An example of H$_2$ cooling gas in a pristine halo core is shown in Figure~\ref{fig:pop3_stages}, where the hydrogen number density reaches $10^3\,\mathrm{cm^{-3}}$ with a temperature of $\sim 300$~K for the {\bf Efficient SF} run. The {\bf HN + High $\epsilon_{\rm ff}$} forms Pop~III stars in generally less dense environments than the other runs, leading to a decrease in Pop~III forming halo masses at later times, as the requirements for cooling efficiency are less strict to achieve star formation densities (see Figure~\ref{fig:pop3_masstime_SFR}).

\begin{figure}
	\includegraphics[width=0.47\textwidth]{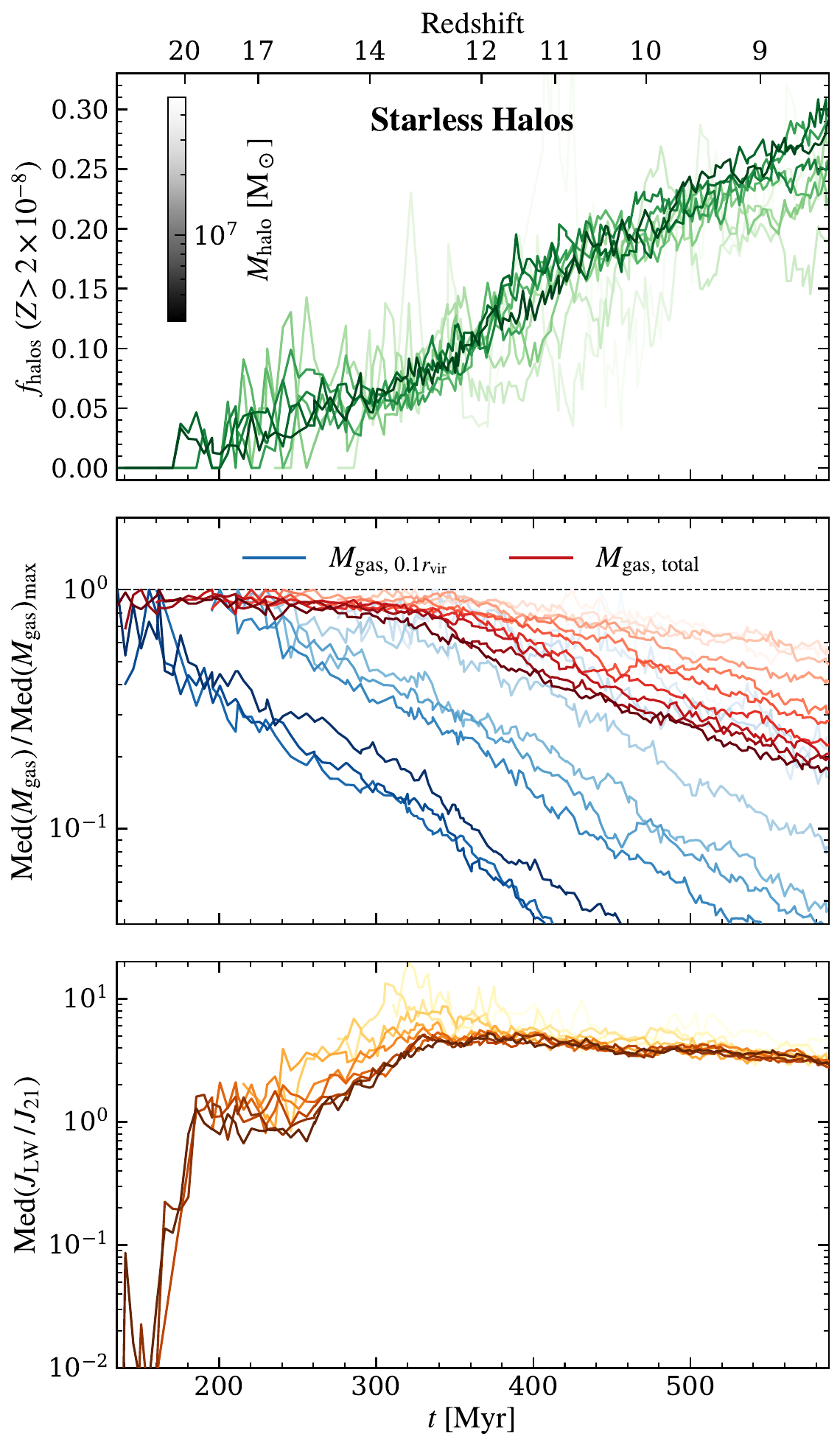}
    \caption{
        \textbf{(Top)} The fraction of halos containing no stars within their virial radii in the {\bf Bursty SF} simulation which have density-weighted metallicities $Z > 2\times10^{-8}$ (the threshold for Pop. II formation) as a function of time. Lines are coloured by halo mass. Starless and metal enriched halos become more common over time, and make up $\sim 20-30\%$ of all starless halos at 600\,Myr. \textbf{(Middle)} The median gas mass of starless halos at a given time divided by the maximum value reached at any time. Red represents the total gas mass in halos out to the virial radius, and in blue indicates the inner gas mass out to $0.1\,r_{\rm vir}$. \textbf{(Bottom)} The median Lyman Werner intensity across starless halos, as measured at the virial radius, as a function of time. The LW intensity increases rapidly after the first stars form, reaching $J_{21}$ at 200~Myr, which suppresses Pop~III formation in mini-halos.
    }
    \label{fig:other_pop3_suppresion}
\end{figure}

LW radiation from neighbouring star forming regions dissociates H$_2$ in low-mass halos, as LW has little opacity in neutral H/He and can penetrate deeply into dense atomic gas residing in pristine halo cores. In our work, we find cooling reduction from LW radiation is the dominant player in halting Pop~III formation in $\sim~4\times10^6\rm~M_\odot$ halos. Within 200~Myr after the Big Bang, Pop~III formation transitions to more massive halos (see Figure~\ref{fig:pop3_masstime_SFR}), whose virial temperatures are $\sim10^4$~K. This critical time ($z\approx 20$) corresponds to the LW background reaching $J_\mathrm{LW} = J_{21} \equiv 10^{-21}\,\rm erg\,s^{-1}\,cm^{-2}\,Hz^{-1}\,sr^{-1}$, as measured at the virial radii of every starless halo in the simulations (see the bottom plot in Figure~\ref{fig:other_pop3_suppresion}, \correction{where we show the strength of the LW field as a function of time for bins of halo masses}).
\correction{
    The LW intensity from local sources in our region is expected to dominate over any a cosmic LW background originating from more distant galaxies, as our zoom-in region is overdense and biased towards early galaxy formation.
}

\begin{figure}
	\includegraphics[width=0.47\textwidth]{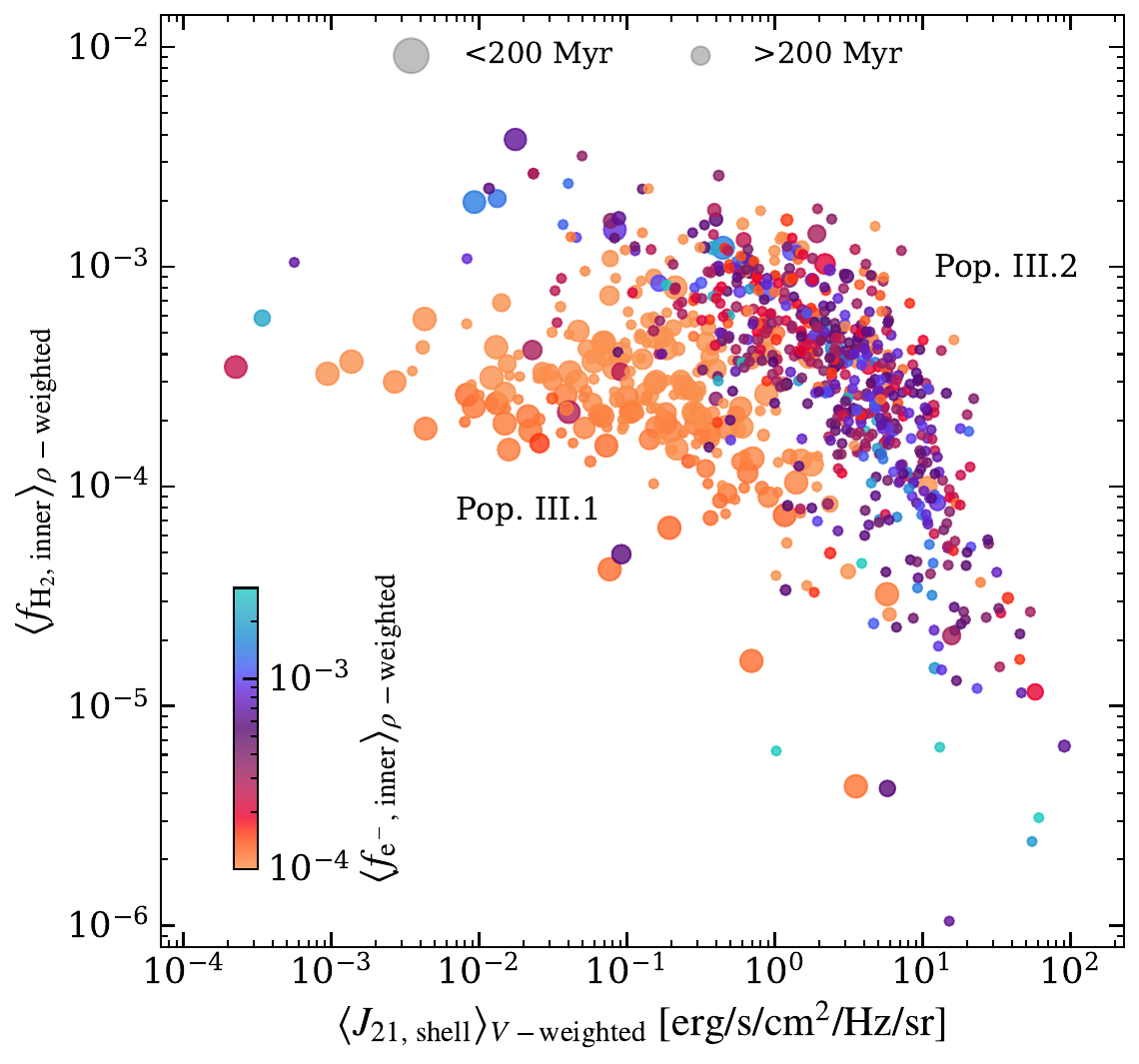}
    \caption{
            The fraction of molecular hydrogen $f_{\rm H_2}$ in the inner $0.1 r_{\rm vir}$ of halos undergoing their first Pop~III formation (within $\sim5$~Myr), as a function of LW intensity (parametrized as $J_{21}$) in a thin shell at the virial radius of the halo, for the {\bf Efficient SF} simulation. Halos are coloured by the electron fraction $f_e$ in the inner $0.1 r_{\rm vir}$, and the sizes of points correspond to the age of Universe at that time. Quantities which are computed in the inner halo are density weighted, while quantities computed in the shell are volume weighted. Two distinct populations emerge: halos with low electron fractions ($f_e \eqsim 10^{-4}$, the Pop~III.1 case) forming at early times, and halos with elevated electron fractions ($f_e > 10^{-4}$, the Pop~III.2 case) forming at later times, across a wide range of LW strengths.
    }
    \label{fig:pop3_forming_fH2_t_fe}
\end{figure}

We show in Figure~\ref{fig:pop3_forming_fH2_t_fe} the H$_2$ fraction in Pop~III-forming halo cores (gas weighted by density in the inner 10\% of the virial radius) as a function of $J_\mathrm{LW}$. Halos are coloured by the electron fraction $f_{e^-} \equiv n_e / n_{\rm H}$ in the dense core. Across all simulations, two distinct populations emerge: halos subject to low levels of LW radiation (with low $f_{e^-} \lesssim 10^{-4}$, the residual electron fraction following recombination, \citealt{Galli_Palla_1998}) at early times and halos able to withstand higher amounts of LW radiation (with high $f_{e^-} > 10^{-4}$) at later times. Halos subject to small amounts of LW radiation are around $\sim3\times10^6\,M_\odot$ at $z\sim25$, while more ionized halos are an order of magnitude more massive and form later on. These two populations arise in the Pop~III.1 and Pop~III.2 formation scenarios respectively \citep{Bromm_2009,Klessen_Pop3Review_2023}.

We find that mini-halos with incoming LW radiation fluxes $J_\mathrm{LW} / J_{21} \geq 1$ do not form Pop~III stars efficiently. Indeed, a strong enough LW field  will dissociate H$_2$ and prevent the gas from cooling, and halos early on are not massive enough to atomically cool. These mini-halos continue to accrete gas, and begin to efficiently cool through atomic hydrogen transitions once their virial temperatures reach $\sim 10^4\,\rm K$ (see the relation to halo mass in Figure~\ref{fig:pop3_masstime_SFR}). Pop~III stars can form in atomic cooling systems that are irradiated by fluxes up to $10^{2}\,J_{21}$. This is due to the extra electrons available in a warmer and more ionized medium, facilitating the formation of H$_2$ through the following pathway \citep{Glover_Federrath_MacLow_Klessen_2010}:
\begin{equation}
    \begin{aligned}
        \mathrm{H} + e^- &\rightarrow \mathrm{H}^- + \gamma, \\
        \mathrm{H}^- + \mathrm{H} &\rightarrow \mathrm{H}_2 + e^-.
    \end{aligned}
    \label{eq:H2}
\end{equation}
Pop~III.2 stars in our simulations mostly form under LW strengths of $0.3 - 30\,J_{21}$, which is a similar range to that seen for Pop~III forming galaxies in the Renaissance simulations at $z=7.6$ \citep{Xu_pop3_2016}. Furthermore, for a given LW strength, more massive halos have consistently larger $f_{\rm H_2}$, as H$_2$ self-shielding becomes more effective due to increased gas masses. Stars in the {\bf HN + High $\epsilon_\mathrm{ff}$} simulation form in less dense gas than the other runs, leading to higher electron fractions from the lack of effective self-shielding.
If the incident LW flux becomes too large, pristine atomic cooling halos could instead proceed to form a direct collapse black hole (DCBH; \citealt{Visbal_2014}), with the spectral shape of the incoming radiation being potentially important to set the H$_2$ and H$^-$ destruction rates \citep{Agarwal_2016}.

Most of the Pop~III-forming halos across all runs have $10^{-5} < f_{\rm H_2} < 10^{-3}$, but there is also a significant population (13-25\% depending on the simulation) with $f_{\rm H_2} < 10^{-8}$ and $J_{21} > 2$. These halos are not shown in Figure~\ref{fig:pop3_forming_fH2_t_fe}  as they are already hosting Pop~III stars, and although the LW intensity at their virial radii is $\sim10\,J_{21}$, it can reach upwards of $10^5\,J_{21}$ in the inner regions of the halo. Such systems have elevated levels of free electrons ($f_e > 2\times10^{-3}$), and are hence able to form H$_2$ on very short timescales after the hosted Pop~III star leaves the main sequence, leading to rapid cooling and the formation of another Pop~III star. This process occurs if radiation pressure is unable to blow out the gas, and implies the former Pop~III star has not undergone a SN, which would enrich and disrupt the gas in the galaxy. \correction{We do not model Ly$\alpha$ radiation pressure (RP), which is an important process that impacts the ISM and would impact efficacy of subsequent SF events or supernova \citep{Nebrin_2025}, as the resonant scattering of Ly$\alpha$ photons could drive significant momentum transfer in dense, dust-free gas \citep{Kimm_2018}. We have run calibration simulations for \textsc{megatron} \citep{katz_impact_2024}, including Pop~III stars and Ly$\alpha$ RP, which does indeed lower the total stellar mass by disrupting star forming regions in low metallicity gas.} We discuss in Section~\ref{subseq: pop3 starbursts} how repeated episodes of Pop~III star formation in the same halo (chained one after another) can lead to Pop~III starbursts in $10^8\,\rm M_\odot$ halos.

To some extent, the Pop~III formation phase can be bypassed if halos are externally enriched with metals from nearby galaxies \citep{Smith_Wise_OShea_Norman_Khochfar_2015,Brauer_AEOS_2025}. In the upper panel of Figure~\ref{fig:other_pop3_suppresion}, we plot the fraction of metal enriched halos hosting no stars as a function of time in the \textbf{Bursty SF} simulation. We find at the latest stages of the simulations, $\sim 20-30\%$ of starless halos in the mass range of $3\times10^6 - 3\times10^7 \;\rm M_\odot$ (the masses of halos which predominantly form Pop~III stars) have elevated density-weighted metallicities of $Z > 10^{-8}$, which is the critical metallicity threshold for Pop.~II formation. As most halos in this mass range are not significantly enriched, we determine Pop~III suppression from metal-enriched outflows of nearby galaxies to be sub-dominant compared to radiative suppression. The number of externally enriched halos in the \textbf{Bursty SF} simulation is an upper bound, as it features the most external metal enrichment of all four simulations, due to the increased supernova energy of Pop.~II stars blowing out more metal mass beyond $r_\mathrm{vir}$ of their host halo.

The decreased possibility for halos to form Pop~III stars could also be linked to the lowering of gas fractions in halos. As FUV radiation increasingly permeates the simulation volume, it photoheats gas in the IGM, pressurizing it and smoothing out small scale fluctuations in its structure. As a result, the funnelling of gas from filaments into DM halos becomes increasingly inefficient as the global star formation rate increases \citep{Dijkstra_Haiman_Rees_Weinberg_2004,Katz_DwarfQuenching_2020}. In a MW-like environment, this process should occur much faster than in an average patch of the IGM \citep{Ocvirk_2016}. 
\correction{
In addition, the IGM could be significantly heated (and ionized) by X-ray sources \citep{Hegde_Furlanetto_2023}. Such heating would further suppress gas accretion onto low-mass halos. Conversely, a higher electron fraction would increase the H$_2$ production rate, so it is unclear whether an X-ray background would help or prevent Pop III formation. We do not explicitly follow X-ray photons, and the uncertainties in the abundance of high-$z$ X-ray sources and their accretion luminosities makes quantifying this effect difficult.
}

In the middle panel of Figure~\ref{fig:other_pop3_suppresion}, we plot the average gas mass in halos containing no stars as a function of time, divided by the maximum average value reached (typically reached in the first several time bins, as the trends are almost monotonous.) We do this for both the total amount of gas in halos (in red), and for the amount of gas within $0.1r_\mathrm{vir}$ (in blue). We find that $\sim 10^6\,\rm M_\odot$ halos forming at 600~Myr have on average a factor of 3-10 (depending on the simulation) times less total gas mass than halos of the same mass forming at 100~Myr. Comparatively, the inner gas reservoirs in the lowest mass halos see an order of magnitude drop over 100~Myr at early times, which is attributed to the photo-dissociation of H$_2$ by LW radiation discussed earlier. The suppression of Pop~III formation at early times is therefore completely dominated by the inability of halos to condense gas into their cores.

Although we capture the diversity of Pop~III environments which would influence their formation properties (Pop~III.1 vs Pop~III.2, see Figure~\ref{fig:pop3_forming_fH2_t_fe}), we do not differentiate between the two populations in our star formation prescription, and instead use a single IMF for both formation scenarios. Pop~III.2 stars could span a lower mass range from effective HD cooling \citep{Nishijima_Hirano_Umeda_2024} leading to higher degrees of freedom for gas fragmentation. We also \correction{do not} consider the Pop~III.1 theory for supermassive BH-progenitor stars ($\sim 10^5\,\rm M_\odot$), driven by dark matter annihilation heating from weakly interacting massive particles \citep[see ][]{Tan_2024}. 

\subsubsection{During the Pop~III Main-Sequence}\label{subsubseq: pop3 during MS}

For haloes where the gas does cool well below the virial temperature, star formation proceeds as described in Section~\ref{subseq: pop3 formation}. Pop~III stars generally have higher luminosities and very hard spectra compared to Pop.~II counterparts of similar mass \citep{schaerer_pop3sed_2002}. The top-heavy nature of our Pop~III IMF means Pop~III stars in \textsc{megatron} primarily produce ionizing photons. The radiation from Pop~III SEDs quickly dissociates the surrounding H$_2$ and photo-heats the dense gas to $10^4$~K, as seen in the middle row of Figure~\ref{fig:pop3_stages}. For the most massive ($>300\,\rm M_\odot$) Pop~III stars, photoheating and radiation pressure can completely blow the gas away. As expected for a hard ionizing source, the spectra feature prominent He\,\textsc{ii} recombination lines which are likely to be unique spectroscopic signatures of Pop~III systems \citep{katz_pop3_2023, Trussler_2023}. During the Pop~III MS, gas in the inner halo is irradiated by LW radiation with intensity in the range $10^3-10^6\,J_{21}$ which maintains the H$_2$ fraction below $10^{-7}$. As expected, the local LW intensity is positively correlated with Pop~III stellar mass and negatively correlated with H$_2$ fraction.

\begin{figure}
	\includegraphics[width=0.47\textwidth]{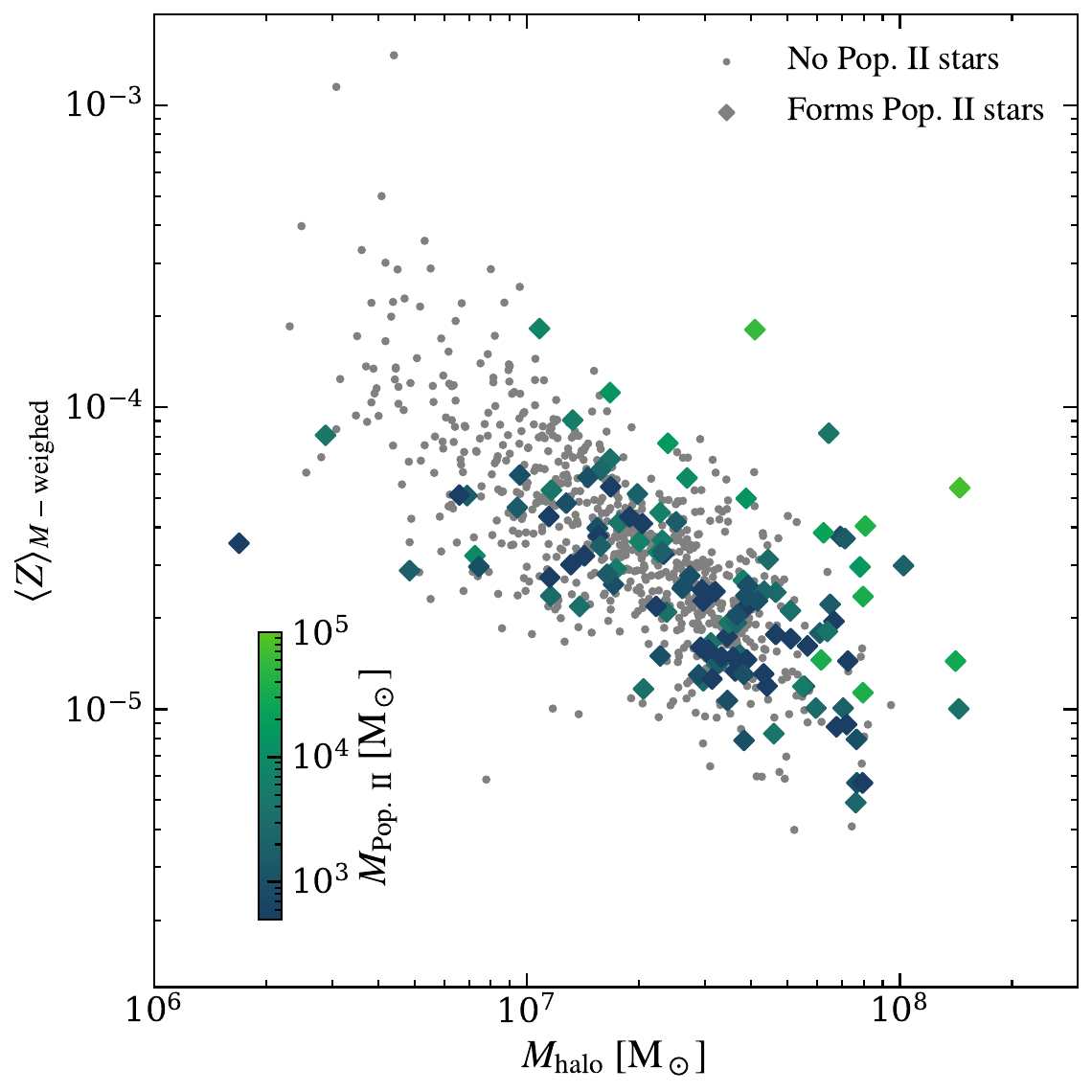}
    \caption{
        The gas phase metallicity in halos as a function of halo mass, $\sim10$~Myr following a PISN from a Pop~III star in the {\bf HN + High $\epsilon_\mathrm{ff}$} simulation. Gray circles are halos that have not yet formed new stars and diamonds represent halos undergoing Pop.~II formation, coloured by their current Pop.~II stellar mass. In general, the metallicity decreases for increasing halo gas mass, as metals are diluted in larger gas reservoirs. Halos with the highest gas masses can grow over $10^5\,\rm M_\odot$ in stellar mass in a short period of time following the PISN.
    }
    \label{fig:pop3_PISN_metallicity}
\end{figure}

\subsubsection{After the Pop~III star dies}\label{subsubseq: pop3 after MS}

The two primary channels of post main-sequence evolution of Pop~III stars in our simulations are either direct collapse into a BH ($\sim\,63\%$) or a PISN explosion ($\sim36\,\%$). If the star forms a BH, it releases no energy or metals\footnote{We do not model subsequent feedback, including RT from BHs which may be accreting.}, allowing the gas to remain pristine. As stated previously, since the gas is no longer subject to a strong radiation field, it quickly reforms H$_2$ and cools again to form another Pop~III star. This process can repeat several times in the same halo core before, statistically, a Pop~III star forms within the mass range for PISN.
If the star undergoes a PISN, upwards of $10^{53}\,\mathrm{erg}$ of energy is released into the surrounding dense gas, transferring momentum which can push significant amounts of gas out of the core.
Along with energy, a PISN also releases metals into the gas, which allows the gas to quickly cool (as seen in Figure~\ref{fig:pop3_stages}) and begin forming the next generation of stars (the Pop. II phase\footnote{See \citealt{Choustikov2025MegP1} for details on the ISM of Pop.~II galaxies in the \textsc{megatron} high-$z$ suite.}). The spectra of such systems show prominent [O\,\textsc{ii}] and [S\,\textsc{ii}] collisional lines, and [O\,\textsc{iii}] if the gas is still hot enough, but are extremely faint due to the lack of stellar radiation. We do find a very small subset of PISN systems which are still hosting a shining Pop~III star.

\begin{figure*}
	\includegraphics[width=0.98\textwidth]{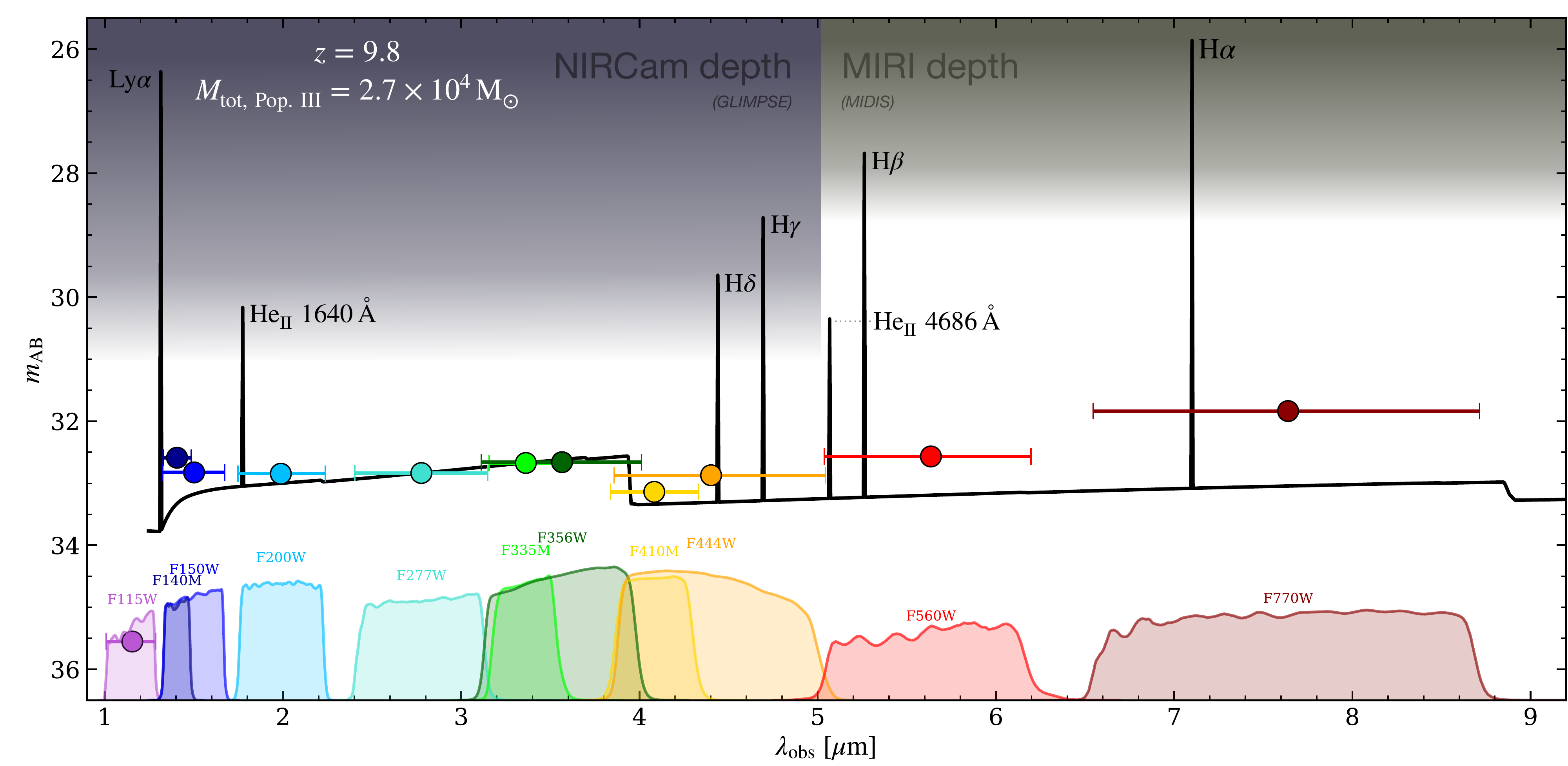}
    \caption{
        AB magnitude of the brightest Pop~III galaxy spectra in our simulations as a function of observed wavelength. The galaxy contains 136 shining Pop~III stars at $z=9.8$ for a combined stellar mass of $2.7\times 10^4\,\mathrm{M_\odot}$. We include mock photometric measurements from the JWST NIRCam and MIRI instruments, with the coloured circles corresponding to a select number of filters. The transmission and range of each filter are shown at the bottom of the figure. The NIRCAM and MIRI depths are illustrated as gradients terminating at $m_\mathrm{AB} \sim 31$ (GLIMPSE; \citealt{Kokorev_2025}) and $m_\mathrm{AB} = 28.75$ (MIDIS; \citealt{Ostlin_2025}), respectively, representing the deepest integration depths reached by those instruments. The spectra feature strong H~$\textsc{I}$ and He~$\textsc{II}$ recombination lines with equivalent widths EW(H$\alpha$)~=~3870~\AA \,and EW($\mathrm{He}_{\textsc{ii},\,1640}$;~$\mathrm{He}_{\textsc{ii},\,4686}$)~=~66~\AA.
    }\label{fig:pop3_spectra_jwstfilters}
\end{figure*}

We find that for all but the smallest halos ($M_\mathrm{halo} \leq 4\times10^6\,\rm M_\odot$) forming Pop~III stars in the PISN mass range, the resulting supernova does not eject most of the gas (including metals) from the halo virial radius, checking the gas phase metallicity content both within 5~Myr and 10~Myr after the supernova.  We show in Figure~\ref{fig:pop3_PISN_metallicity} the mass-weighted metallicity as a function of halo mass following a PISN event.
In general, larger halos contain more gas. As a result, the metallicity decreases with increasing gas mass as metals are diluted in larger gas reservoirs. The scatter in metallicity for a given halo mass in Figure~\ref{fig:pop3_PISN_metallicity} is partly driven by the mass range of Pop~III stars undergoing PISN, which span a factor of a few in metal yields depending on the initial mass of the star.
The gas phase metallicity in halos are all greater than $10^{-6}$, which is enough enrichment to begin the Pop.~II formation phase.
After $\sim10$~Myr we find that $45\%$ of halos that hosted a PISN are now forming Pop.~II stars in the {\bf Efficient SF} simulation, compared to only $10\%$ in the {\bf HN + High $\epsilon_\mathrm{ff}$} where environments are typically over an order of magnitude less dense. Although metals and gas are retained by halos, we may not be fully resolving the complex mixing of metals following a PISN, as we find the majority of metals to efficiently mix into the cold phase of the ISM on a $\sim10$~Myr timescale, quickly cooling the enriched gas and driving the beginning of the Pop.~II star formation phase.

Our results differ from \citet{Mead_AEOS_2025}, who find the majority of mini-halos to expel all metals following supernovae, and their Pop~III stars are much smaller and only include CCSNe with $10^{51}\,$erg. Since halos forming Pop~III stars in the AEOS simulations are an order of magnitude less massive than those in \textsc{megatron}, metals and indeed the rest of the gas is retained by the steeper gravitational potential \citep{Kitayama_Yoshida_2005,Whalen_2008,Cooke_Madau_2014}. In fact, halos with the lowest gas masses ($M_\mathrm{gas}\approx 10^5\rm~M_\odot$) can reach metallicities of $6\times10^{-3}$ following a PISN, as metals are diluted less. As a result of PISN enrichment, there exists an iron metallicity plateau at $\rm [Fe/H] \approx -2.5$ in the gas phase of small dwarf galaxies \citep[see][for more details]{Rey2025MegP1}.

\subsection{Pop~III starbursts in massive galaxies}\label{subseq: pop3 starbursts}

We find in all simulations a few Pop~III starbursts occurring in halos up to $3\times10^8\,\rm M_\odot$. The starbursts are comprised of around 20-130 Pop~III stars forming in rapid succession, with some visible as outliers in Figure~\ref{fig:pop3_masstime_SFR}. As these systems have the largest amount of Pop~III stars shining at the same time, they represent the brightest Pop~III galaxies in our simulations The total gas masses in Pop~III starburst halos range from $2\times10^7 - 4\times10^7\,\rm M_\odot$, and are found to have been the sites of previous Pop~III formation.

~\cite{Xu_pop3_2016} are also able to form Pop~III starbursts in their halos, although in their simulation they form in halos with lower gas masses of $6\times10^6$ to $10^{7}\,\rm M_\odot$. While they attribute this growth to external LW radiation preventing collapse in the vicinity of bright galaxies, we find our halos to self-quench cooling through the repeated formation of Pop~III stars in the BH mass range. This allows for the host halo to continue growing while keeping the gas pristine, until a critical gas mass is reached to trigger a Pop~III starburst. Since the halo has built up a large reservoir of dense gas, Pop~III stars can keep forming even under the strong LW radiation and photoheating sourced from the stars.

Our results for Pop~III starburst galaxies could change significantly if assuming a different Pop~III IMF, or if the thermochemistry was better calibrated for the Pop~III.2 formation case. As Pop~III.2 stars could be potentially formed from a less top-heavy IMF, a lower mass range would yield fewer black holes and therefore a smaller probability for Pop~III starbursts. 
\correction{
The rapid star formation process could also be halted sooner if Ly$\alpha$ RP was included in our simulations (see the calibration simulations including Ly$\alpha$ RP in \citealt{katz_impact_2024}). In metal poor gas, with stars that produce many ionizing photons, Ly$\alpha$ becomes a dominant source of pressure \citep{Nebrin_2025}. Such pressure would potentially drive gas out of the collapsing cloud and prevent further star formation, although the detailed structure of the ionized region needs to be resolved to properly quantify this effect.
}

While our simulations are able to produce Pop~III starbursts, we stress that they are extremely rare events (1-3 per simulation) and not representative of the larger population of Pop~III systems. Nevertheless, they are fundamentally interesting objects as they are the easiest to observe.

\subsection{The Observability of Pop~III galaxies}

\subsubsection{Spectra of a bright Pop~III galaxy}\label{subseq: pop3 spectra}

The spectra of Pop~III systems in \textsc{megatron} are diverse, with UV slopes in the range $-2<\beta<-1.5$ (redder than those of typical Pop.~II galaxies, see \citealt{Katz2025MegP1,Katz2025MegP2}). Equivalent widths of the He \textsc{ii} lines depend strongly on the mass of the Pop~III stars, and typically range from 10-200~\AA. Equivalent widths of the H$\alpha$ line also vary, with values between 3000-5000~\AA. The top-heavy nature of our Pop~III IMF used in the simulations means that these stars have very efficient production of ionizing photons, leading to nebular emission brighter than the stellar spectra itself \citep{Cameron_2024, Katz_2025}. Here, we focus on the spectra of the brightest Pop~III galaxy produced in the \textsc{megatron} simulations, which we show in Figure~\ref{fig:pop3_spectra_jwstfilters} as $m_\mathrm{AB}$ against observed wavelength $\lambda_\mathrm{obs}$. We convolve the spectra with JWST NIRCam and MIRI filters as they represent the deepest existing observations.

\begin{figure}
	\includegraphics[width=0.47\textwidth]{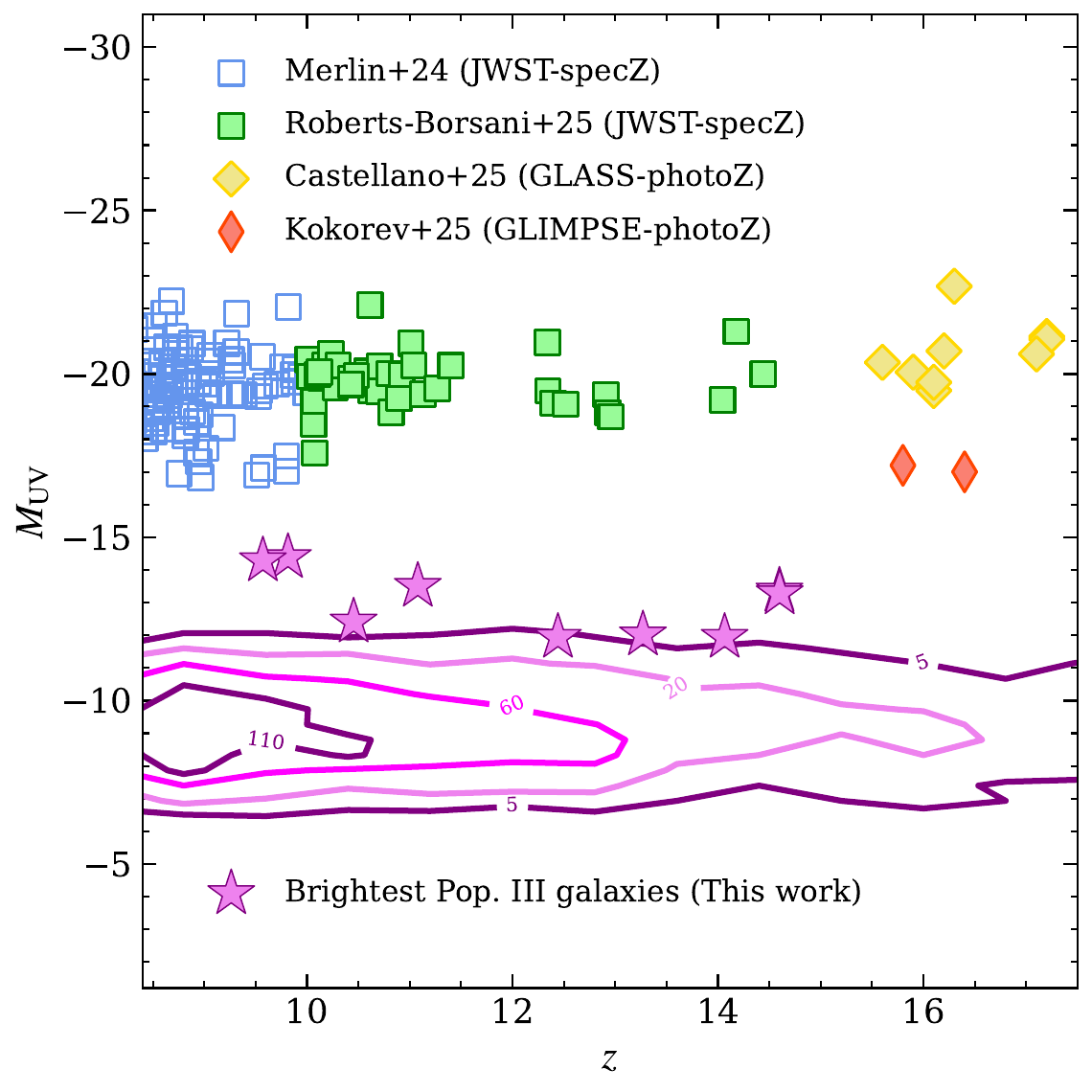}
    \caption{
        The UV magnitudes of the 10 brightest Pop~III galaxies in \textsc{megatron} as a function of redshift across all the simulations. All other Pop~III galaxies are represented by density isocontours with the count thresholds as inline labels. We include the highest redshift galaxies observed by JWST, whose redshift has either been spectroscopically confirmed \citep{Merlin_2024,Roberts-Borsani_2025}, or inferred from NIRCam photometry \citep{Castellano_2023, Kokorev_2025}. Pop~III galaxies generally have $-7 > M_\mathrm{UV} > -12$, and the brightest galaxy reaches $M_\mathrm{UV} =- 14.5$ which is still 2 dex below that of the faintest JWST galaxy observed.
    }\label{fig:pop3Muv_vs_JWST_galaxies}
\end{figure}

The strongest emission lines in the spectra are Ly$\alpha$ and H$\alpha$, the first spectral lines in the Lyman and Balmer series respectively. Whilst a booming H$\alpha$ line (with equivalent widths greater than 3000) is an indicator of Pop~III systems, the Ly$\alpha$ line is likely to be heavily attenuated by the neutral IGM in the epoch of reionization \citep{Inoue_2014}. However, since galaxies both make and reside in ionized bubbles, the Ly$\alpha$ emission from Pop~III galaxies might be detectable, as we have already observed the Ly$\alpha$ transition in objects beyond redshift 10 \citep{Bunker_2023, Witstok_2025}. The size of the bubble depends on redshift, neutral fraction of the IGM, and EW of the Ly$\alpha$ line. It is unclear whether such a galaxy would have enough ionizing flux to carve out an ionized bubble out to a few pMpc, required to redshift Ly$\alpha$ out of the resonance wavelength \citep{Saxena_2023}, and would probably need neighbouring systems to help re-ionize the IGM. The equivalent width of H$\alpha$ of our brightest Pop~III galaxy is measured at EW(H$\alpha$)~=~3870~\AA, and both $\rm He\,\textsc{ii}$ lines have equivalents widths of EW($\mathrm{He}_{\textsc{ii},\,1640}$;~$\mathrm{He}_{\textsc{ii},\,4686}$)~=~66~\AA. The recently discovered system LAP2-b \citep{Vanzella_2025} has a derived stellar mass of at most $2\times10^4\,\rm M_\odot$ which is comparable to our brightest Pop~III galaxy, but has a measured EW($\mathrm{H}\alpha$) = 647~\AA\ which is $6-8$ times weaker than our Pop~III galaxies.

The burden of proof for Pop~III systems is incredibly high, as it involves proving the absence of metals. Photometric and spectroscopic observations only yield upper limits on metal lines, and we would need to measure the nebular continuum in exquisite detail to confirm the absence of carbon lines and rule out a high enough carbon abundance for the CNO cycle to operate. As the [O\,\textsc{iii}] 5007~\AA{} line is the most prominent metal line accessible at high redshift, a potentially important diagnostic for constraining spectra into the Pop~III regime is the [O\,\textsc{iii}]\,/\,H$\beta$ ratio \citep{katz_pop3_2023}. For almost all Pop~III galaxies in \textsc{megatron}, the spectra are nebular-dominated and only feature recombination lines of H and He. We do however find rare cases where multiple Pop~III stars forming at the same time, with one star going PISN while the others remain on the MS, leading to bright [O\,\textsc{iii}], [O\,\textsc{ii}], and [S\,\textsc{ii}] emission.

\begin{figure}
	\includegraphics[width=0.47\textwidth]{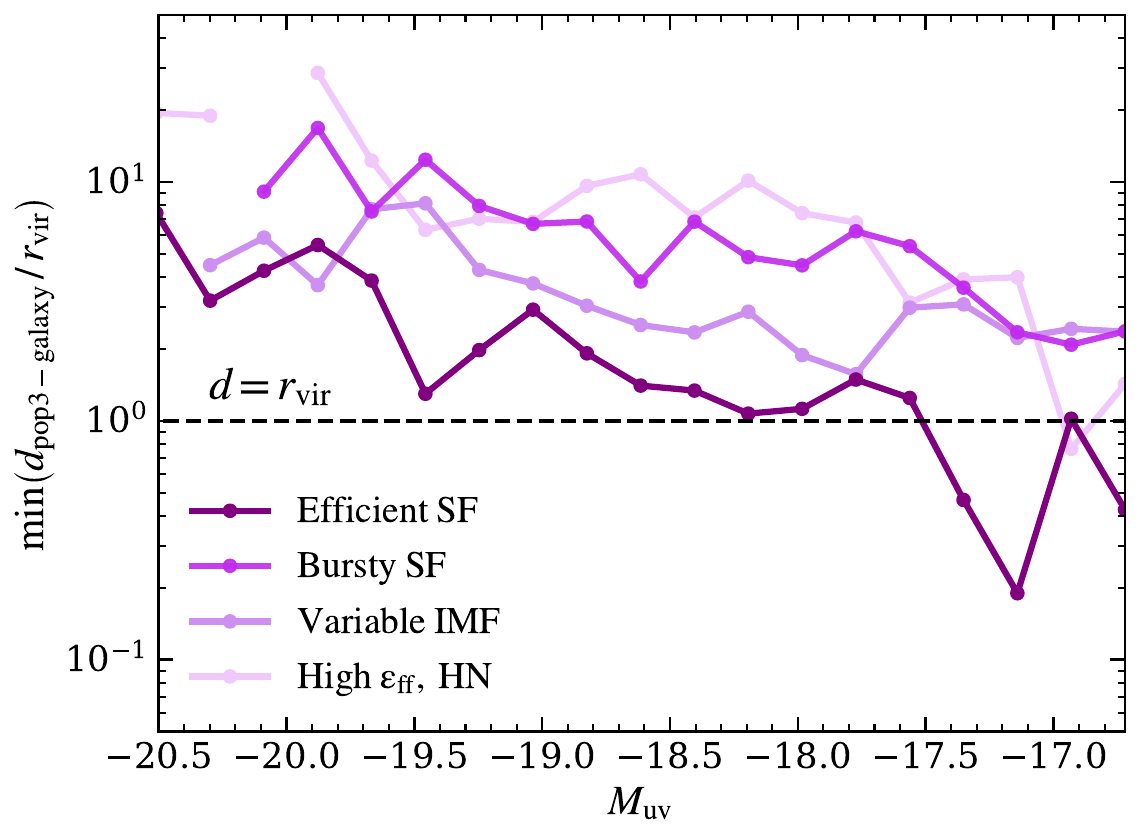}
    \caption{
        The minimum distance between a recently formed Pop~III star and UV-bright galaxies, divided by those galaxies' virial radii, as a function of $M_{\rm UV}$. The coloured lines are for the four different \textsc{megatron} simulations. Pop~III stars are able to form closer to UV-bright galaxies for increasing $M_{\rm UV}$, with the {\bf Efficient SF} run forming Pop~III stars within $r_{\mathrm{vir}}$ of JWST-observable galaxies beyond $z=8$.
    }\label{fig:pop3_to_UVbright_r_rvir_ratio}
\end{figure}

\begin{figure*}
	\includegraphics[width=0.99\textwidth]{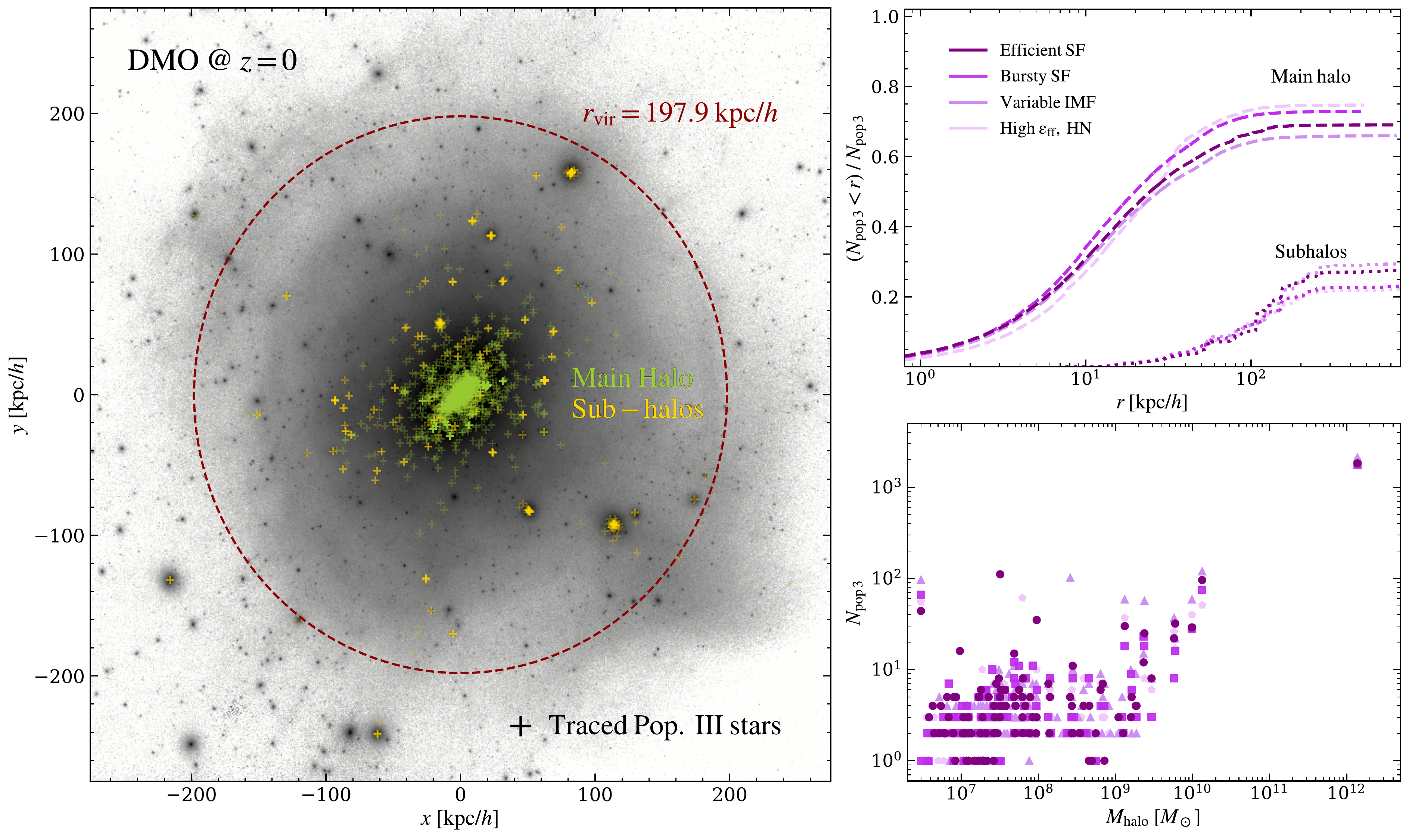}
    \caption{
    \textbf{(Left)} Particle projection of the dark matter structure in the dark matter-only (DMO) simulation at $z=0$, centred on the main halo. The virial radius of the halo is indicated by the circle drawn with a dark red dashed line. Crosses represent a Pop~III star traced down to $z=0$ by following the closest 25 dark matter particles in the last output of the \textbf{Efficient SF} simulation. We pick a random dark matter particle from the list of 25, plot a cross at its position at $z=0$, and colour the cross in green if it is in the main halo, or in yellow if it is in a sub-halo. \textbf{(Upper Right)} The CDF of the distances between the Pop~III stars and the centre of the main halo at $z=0$, for both main halo and sub-halo stars. \textbf{(Lower Right)} The number of Pop~III stars in each halo as a function of halo mass, coloured by which simulation do the Pop~III stars originate from.
    }\label{fig:pop3_tracing_DMO}
\end{figure*} 

The physical resolution at the onset of Pop~III formation (1.5~pc), combined with the top-heavy IMF, means that almost all Strömgren spheres are resolved at $z=30$. As the simulation proceeds to lower redshift, the expanding Universe coarsens the resolution of our grid. By the end of the simulation ($z=8.5$), most Pop~III stars no longer have resolved Strömgren spheres except for the most massive ones, leading to a less accurate processing of ionizing photons and potentially biasing emission line/nebular continuum predictions (although we apply CLOUDY corrections to cells with unresolved Strömgren spheres\footnote{\textsc{ramses-rtz} has been shown to be within 10\% of CLOUDY for idealized tests \citep{katz_impact_2024}}, see \citealt{Katz2025MegP1} for details.)

\subsubsection{UV magnitudes}\label{subseq: pop3 UV mags}

We compute the UV magnitudes of every halo hosting an actively shining Pop~III star in the \textsc{megatron} simulations using the method described in Section~\ref{subseq: generating spectra}
We plot in Figure~\ref{fig:pop3Muv_vs_JWST_galaxies} the UV magnitudes of our simulated Pop~III galaxies as a function of redshift, and compare them to some of the faintest galaxies detected by JWST. The vast majority of Pop~III galaxies have $-6 > M_\mathrm{UV} > -12$, with the distribution peaking around $M_\mathrm{UV} \sim -9$. The brightest Pop~III galaxy in our simulations reaches $M_{\rm UV}~\sim~-14.5$ at $z=9.8$, and is around two orders of magnitude fainter than the faintest galaxies detected by JWST beyond $z=8$ \citep{Merlin_2024,Roberts-Borsani_2025,Castellano_2023,Kokorev_2025}. A gravitational lens with magnification $\mu > 10$ would allow the Pop~III galaxy spectrum to be detectable by deep NIRCam imaging if considering integration times similar to those of the GLIMPSE survey. While such a scenario might seem unlikely \citep{Vikaeus_2022,Zackrisson_2024}, there are potentially pristine and star-forming systems already being observed at $z\sim 4$ under a strong $(\mu~\sim~50)$ magnifying lens \citep{Vanzella_2025}.

\subsubsection{Proximity to UV-bright galaxies}\label{subseq: pop3 UV galaxies}

Current searches for Pop~III stars using JWST either focus on gravitational caustics, where lensing of a background galaxy can boost the observed flux of individual star forming regions by an order of magnitude, or look at the surrounding environments of UV-bright galaxies. Some of the faintest objects JWST is currently detecting beyond $z=8$ have $M_{\rm UV} \sim -17$ (see Figure~\ref{fig:pop3Muv_vs_JWST_galaxies}), so we adopt this value as our lower limit for UV-bright galaxies. In Figure~\ref{fig:pop3_to_UVbright_r_rvir_ratio}, we show the closest distance between newly formed Pop~III stars and galaxies as a function of $M_{\rm UV}$. In general, the minimum distance between Pop~III stars and UV-bright galaxies decreases with increasing $M_{\rm UV}$. The {\bf Efficient SF} run consistently has Pop~III stars closest to bright galaxies, with some Pop~III stars forming within the virial radius of galaxies with $-17.5 < M_{\rm UV} < -17$. This simulation produces many more UV-bright galaxies than the other three simulations due to it having the weakest feedback. The large number of bright galaxies occupy more of the volume which increases the probability of having nearby Pop~III stars.

Out of the four \textsc{megatron} runs, only the {\bf Efficient SF} and {\bf HN + High $\epsilon_{\rm ff}$} simulations are able to produce Pop~III stars within the virial radius of bright galaxies. Consistent across all runs, the vast majority of Pop~III stars form at least $10r_{\rm vir}$ away from bright galaxies, with only $6/\sim10^{4}$ Pop~III stars forming within the virial radius of UV-bright galaxies, challenging the feasibility of detecting Pop~III stars in this manner. As we find a general trend of Pop~III stars forming closer to increasingly dimmer galaxies, and if JWST is able to detect galaxies further down to $M_{\rm UV} = -15$ (perhaps even down to $M_{\rm UV} = -12$, see \citealt{Chemerynska_2025}), then the prospects of finding Pop~III stars in the vicinity of UV-bright galaxies could improve. Whether through proximity of UV-bright galaxies or behind strong gravitational lenses, detecting Pop~III stars at high-$z$ remains an onerous enterprise, and it very well could be that the current approaches of observational surveys are the most optimal.

\subsection{Pop~III distribution at z=0}\label{subseq: pop3 present day}

An interesting open question regarding Population III stars concerns where they end up at the present day. Many surveys have looked for very metal poor stars \citep{Frebel_Norris_2015}, whose chemical abundances are the result of individual Pop~III supernova \citep[e.g.][]{Ji_2020}. If Pop~III stars are able to form below $0.8\,\rm M_\odot$, then they could potentially survive to $z=0$ \citep{Maeder_Meynet_Pop3Review_2012}, and have unique spectroscopic signatures (specifically the lack of metal lines, both in emission and absorption) detectable by upcoming archaeological surveys like 4MOST \citep{Christlieb_2019} and WEAVE \citep{Jin_2024}. As our IMF of choice is very top-heavy, none of the Pop~III stars in \textsc{megatron} have main sequence lifetimes long enough to live to the present day, they instead become BHs or explode as SN. However since the Pop~III stars in our simulations are relics of pristine and dense star-forming gas clouds, regardless of the specific details of the uncertain Pop~III IMF, our Pop~III stars trace the sites of metal-free star formation. Nevertheless, given the uncertainties in the Pop~III IMF, if we temporarily assume that some low mass Pop~III stars can form at these sites, we can ask where these stars might be at the present day.

We track the positions of Pop~III stars to $z=0$ following the method described in Section~\ref{subseq: pop3 traceback method}, and plot their spatial distribution in Figure~\ref{fig:pop3_tracing_DMO}, coloured by whether they reside in main halos or subhalos. We include the number of Pop~III stars per halo as a function of halo mass, and the cumulative distribution function of Pop~III stars as a function of distance from the main halo centre. We find the majority of Pop~III stars reside in the main halo, with 70\% of main halo stars found further than 10~kpc ($>0.02r_\mathrm{vir}$) from the halo centre.\ \citet{Starkenburg_APOSTLE_2017} also find a dominant fraction of Pop~III stars reside in the stellar halos of the APOSTLE simulations. Furthermore, we find 20-25\% of Pop~III stars are in subhalos that have not undergone major tidal disruption. Our results are in close agreement with those of \citet{Yang_Gao_Guo_Li_Shao_Zhao_2025}, who analyse halos from the Auriga project and find a similar 80-20 split between their first stellar population\footnote{These simulations do not explicitly form Pop~III stars, and instead call their first stellar population ``earliest star relic''} in the main halo and in subhalos, respectively. 

Individual subhalos can host upwards of 100 Pop~III remnants, even down to halo masses of $2\times10^6\,\rm M_\odot$. While one might expect to find more Pop~III stars in systematically larger subhalos, the relation is not as clear. The trend remains relatively flat in $N_\mathrm{pop3}$ for halos with masses $<10^9\,\rm M_\odot$, with a steep increase in $N_\mathrm{pop3}$ for halos more massive than $10^9\,\rm M_\odot$. While we \correction{can not} make a determination on the final stellar masses of the subhalos in the DMO simulation, we can assume they are comparable to local dwarf galaxies of similar halo mass. When considering the derived stellar masses of the Milky Way and surrounding satellites, the best regions to search for Pop~III stars may well be in small dwarfs whose $M_{\star, \,\rm Pop.\,III} / M_{\star, \,\rm total}$ ratios are highest \citep{Magg_Hartwig_Agarwal_Frebel_Glover_Griffen_Klessen_2018,Klessen_Pop3Review_2023}.

\section{Conclusions}\label{seq:Conclusions}

We have analysed the \textsc{megatron} suite of simulations of a Milky Way progenitor, run with the \textsc{ramses-rtz} code \citep{katz_ramses-rtz_2022}, to understand the formation of Population III stars in the early Universe, focusing on the environments in which they form, their observational properties, and where their remnants end up at the present day. \textsc{megatron} is well suited to study the formation of the first stars because:
\begin{itemize}[leftmargin=*]
    \item it encompasses \emph{the large scale environment} of a Milky Way-mass progenitor
    \item it follows \emph{a non-equilibrium thermochemical network}, including the formation of H$_2$
    \item it includes \emph{on-the-fly and multi-frequency radiative transfer} from star particles, including the Lyman Werner band which is a crucial component for studying Pop~III environments
    \item it has \emph{high resolution} ($\lesssim \SI{1.5}{pc}/h$) at the onset of the formation of Pop~III stars, resolving the individual Strömgren spheres of the earliest stars
    \item it includes a \emph{dark matter-only version run to $z=0$}, allowing us to connect the remnants of Pop~III stars to their spatial distributions at the present day.
\end{itemize}
We are not the first to model Pop~III formation, nor do we have the best resolution, but we have all five points above. Particularly, the environments of Pop~III stars in a proto-Milky Way has not yet been studied with such prescriptions (although see \citet{Griffen_2018} in the context of semi-analytic models).
The main findings from our analysis are that:
\begin{enumerate}[leftmargin=*]
    \item  Pop~III stars preferentially form in halos in the mass range $\sim 10^6 - 10^8 \;\rm M_\odot$ with a constant SFR of $10^{-3} \;\rm M_\odot \, yr^{-1}$ up to the end of the simulations ($z\approx 8.5$). The presence of Lyman Werner radiation from nearby star forming regions can dissociate H$_2$ and prevent Pop~III formation in $10^6\;\rm M_\odot$ halos, pushing Pop~III star formation to more massive halos ($\sim 10^7 - 10^8 \;\rm M_\odot$). Although we find an increase in starless but metal-enriched halos over time, the majority of starless halos remain pristine down to $z\sim 8.5$.

    \item  On extremely rare occasions, the order of a few per simulation, a halo can host the formation of many Pop~III stars. If the repeated Pop~III star formation events all conclude with direct collapse to a BH, the host halo will continue to grow and remain metal-free, until a critical gas mass of $\sim10^8\,\rm M_\odot$ is reached to trigger a Pop~III starburst. These starbursts can form up to 100 Pop~III stars in rapid succession in halos up to $2\times10^8\,\rm M_\odot$. If halos in the Universe are able to form such starbursts, they would be the most likely to be directly observable.

    \item  Although there is a clear trend of Pop~III stars forming closer to galaxies as a function of increasing $M_{\rm UV}$, Pop~III stars typically form far away from UV-bright galaxies, with only $\sim 6$ out of $10^4$ Pop~III stars forming within the virial radius of galaxies with $M_{\rm UV} < -17$. This suggests that detecting Pop~III stars as a consequence of their proximity to bright galaxies is unlikely, but is probably the most optimal approach compared to a random search.

    \item  Finally, we trace the positions of Pop~III stars to $z=0$ using a dark matter-only version of our simulations and find that most of them (>70\%) reside in the main halo, with a significant fraction (20-25\%) in subhalos. Our conclusions are similar to those of previous studies \citep{Starkenburg_APOSTLE_2017,Yang_Gao_Guo_Li_Shao_Zhao_2025}, which also found that Pop~III stars are predominantly located in the diffuse stellar halos of their host galaxies. Additionally, there are tidally undisturbed halos down to masses of $10^6\,\rm M_\odot$ which can host up to $\sim 100$ Pop~III stellar remnants
\end{enumerate}

In recent years, we have become increasingly reliant on numerical simulations to self-consistently model the of environments of Pop~III stars \citep{Xu_pop3_2013, kimm_pop3_2017, katz_pop3_2023, Brauer_AEOS_2025}, particularly the inhomogenous nature of the LW background and its dependence on the overdensity structure sourced in the initial conditions. Our work is the first to self-consistently follow the LW background in the progenitor of a MW-like environment along with modelling the formation of individual Pop~III stars. Although simulating such environments with a high enough resolution to resolve the Pop~III phase remains computationally difficult, the \textsc{megatron} simulations highlight the importance of following the radiation field in such an environment.

The distribution of Pop~III stars in a MW environment at present day have not yet been derived from self-consistent modelling of Pop~III formation. With \textsc{megatron}, we are able to connect the formation of Pop~III stars to the halo-subhalo structure at $z=0$. Our results therefore provide a key theoretical baseline for the search of Pop~III stars in Near-field cosmology. There are many current efforts to find Pop~III stars, with an ever increasing catalogue of metal poor stars in the MW stellar halo \citep{Christlieb_2019} and nearby dwarf galaxies \citep{Skuladottir_2023}. Simulations such as \textsc{megatron} will be key to interpret results for many of the upcoming surveys, allowing to constrain the early evolution of galaxies like our Milky Way.

\section*{Acknowledgements}

AFS would like to thank Santi Roca-Fàbrega for their helpful discussions and insights. AFS and NC acknowledge support from the Science and Technology Facilities Council (STFC) for the PhD studentships. TK is supported by the National Research Foundation of Korea (RS-2022-NR070872 and RS-2025-00516961) and by the Yonsei Fellowship, funded by Lee Youn Jae. OA acknowledges support from the Knut and Alice Wallenberg Foundation, the Swedish Research Council (grant 2019-04659), the Swedish National Space Agency (SNSA Dnr 2023-00164), and the LMK foundation.

This work was performed using the DiRAC Data Intensive service at Leicester, operated by the University of Leicester IT Services, which forms part of the STFC DiRAC HPC Facility (www.dirac.ac.uk). The equipment was funded by BEIS capital funding via STFC capital grants ST/K000373/1 and ST/R002363/1 and STFC DiRAC Operations grant ST/R001014/1.
This work used the DiRAC@Durham facility managed by the Institute for Computational Cosmology on behalf of the STFC DiRAC HPC Facility (www.dirac.ac.uk). The equipment was funded by BEIS capital funding via STFC capital grants ST/P002293/1, ST/R002371/1 and ST/S002502/1, Durham University and STFC operations grant ST/R000832/1.
This work was performed using resources provided by the Cambridge Service for Data Driven Discovery (CSD3) operated by the University of Cambridge Research Computing Service (www.csd3.cam.ac.uk), provided by Dell EMC and Intel using Tier-2 funding from the Engineering and Physical Sciences Research Council (capital grant EP/T022159/1), and DiRAC funding from the Science and Technology Facilities Council (www.dirac.ac.uk). DiRAC is part of the National e-Infrastructure.
This work has made use of the Infinity Cluster hosted by Institut d'Astrophysique de Paris. We thank Stephane Rouberol for running smoothly this cluster for us. The material in this manuscript is based upon work supported by NASA under award No. 80NSSC25K7009.
The authors thank Jonathan Patterson for smoothly running the Glamdring Cluster hosted by the University of Oxford, where part of the data processing was performed.
The authors also acknowledge financial support from Oriel College’s Research Fund.

This project made extensive use of the following packages: \textsc{numpy} \citep{Harris_NUMPY_2020}, \textsc{scipy} \citep{Virtanen_SCIPY_2020}, \textsc{matplotlib} \citep{Hunter_MATPLOTLIB_2007}, \textsc{pandas} \citep{reback_PANDAS_2020}, \textsc{astropy} \citep{Astropy_Collaboration_2022}, and \textsc{yt} \citep{turk_yt_2011}.

\section*{Data Availability}

The data used in this paper can be obtained on reasonable request to the corresponding author.

\bibliographystyle{mnras}
\bibliography{references}

\appendix

\section{Tracing DM particles}\label{app:traceback_validation}

\correction{We show in Figure~\ref{fig:traceback_validation} a validation of our particle traceback method using halos and their constituent DM particles from the DMO simulation. We select the 25 particles which are most bound to subhalos with masses $M_{\rm halo} > 2\times10^7\,\rm M_\odot$ at $z=8.5$, and compute the distribution of distances from those particles to the most bound particle (proximity to halo centre) at that redshift (thin lines). We then track those same particles to $z=0$ and compute the distribution of distances to the same particle again (thick lines). The yellow lines correspond to subhalos which survive as distinct structures until $z=0$, while the green lines correspond to subhalos which merge into the central halo by $z=0$.}

\correction{We find that particles in halos which merge into the central halo have average distances on the order of $\sim10$~kpc at $z=0$, as the particles follow random orbits in the virialized main halo. In contrast, we see that particles which belong to subhalos that survive to $z=0$ remain relatively clustered together, with a minority of particles being stripped away to larger distances. This validates our approach of tracing Pop~III star particles to $z=0$ by following the dark matter particles which were most bound to their host halos at the time of formation.}

\subsection{Minimizing phase-space instead of distance}

\begin{figure}
	\includegraphics[width=0.47\textwidth]{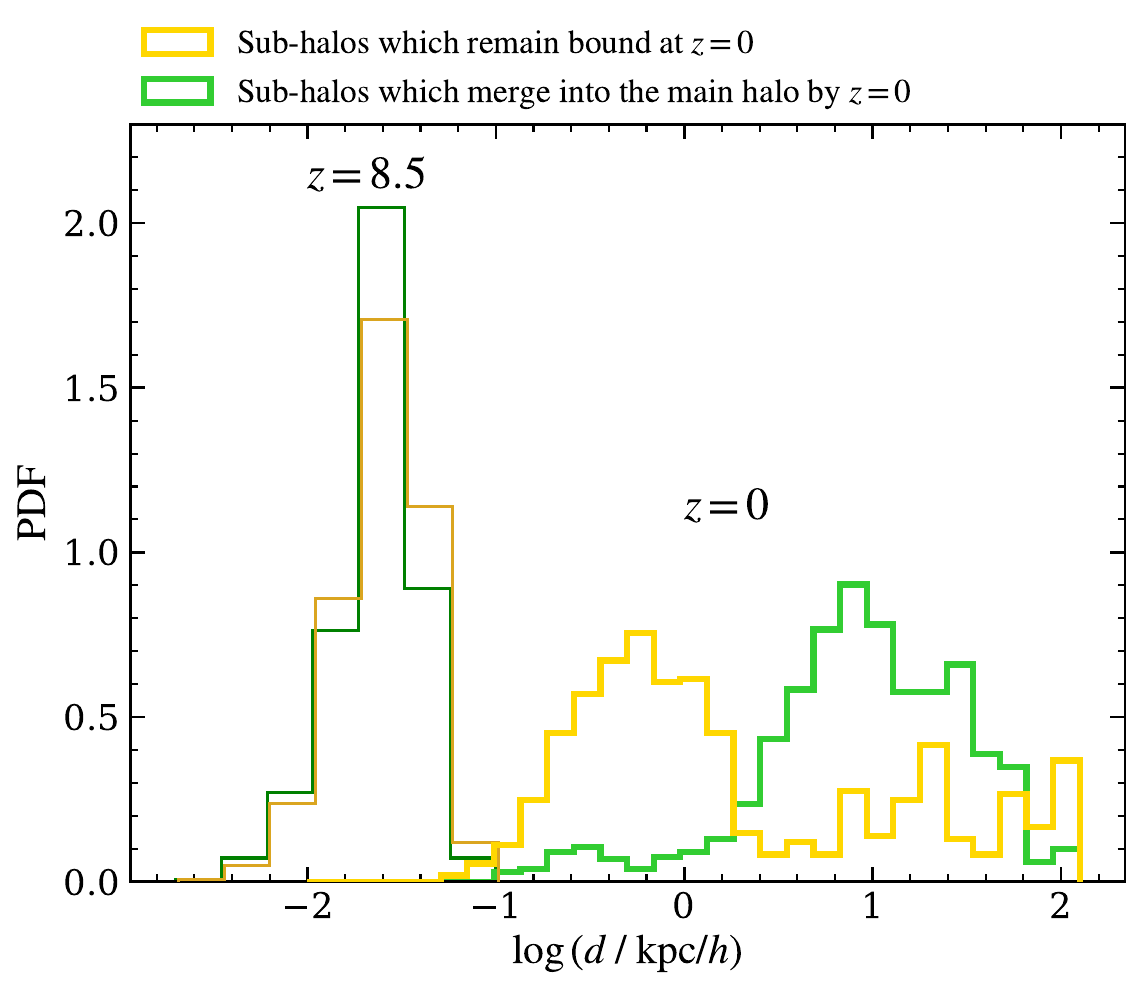}
    \caption{
        \correction{
        Distribution of particle distances to the most bound particle in their halo at $z=8.5$ (thin lines), and the distribution of particle distances to the same particle at $z=0$ (thick lines). We differentiate between subhalos which survive to $z=0$ (yellow), and those which merge into the central halo (green). The distributions are done for 50 halos and stacked. Subhalos which remain bounded at $z=0$ have their particles remain more clustered together, while those which merge see their particles spread out over a larger volume (as the main halo virializes).
        }
    }\label{fig:traceback_validation}
\end{figure}

\correction{
Instead of choosing the 25 DM particles physically closest to the Pop~III star, we also explored using the 25 DM particles which minimized the phase space criteria $d_{\rm phase}$. Following the approach of \citet{behroozi_rockstar_2013}, we define our phase space criteria $d_{\rm phase}$ between two particles $p_1$ and $p_2$ in an ensemble of many points as
\begin{equation}
    d_{\rm phase}(p_1, p_2) = {\left(\frac{|\boldsymbol{x}_1 - \boldsymbol{x}_2|^2}{\sigma_x^2} + \frac{|\boldsymbol{v}_1 - \boldsymbol{v}_2|^2}{\sigma_v^2}\right)}^{1/2},
\end{equation}
where $\boldsymbol{x}$ and $\boldsymbol{v}$ are the position and velocity of the particles, and $\sigma$ is the standard deviation on those quantities across all particles in the system. We found that for a small subset of dark matter particles $(\sim 10)$ near a Pop~III star, choosing them based upon the phase-space criteria was not significantly better at tracking the star from $z=20$ to $z=8.5$, compared to choosing the physically closest particles to the star. We also found that simply increasing the amount of particles sampled around the Pop~III star improved the tracking of the star, more-so than attempting to cleverly choose a subset which matched more closely the kinematics of the star. Furthermore, the $z=0$ distributions in~\ref{fig:traceback_validation} do not meaningfully change when switching to a phase-space criteria for particle selection.
}

\bsp%
\label{lastpage}
\end{document}